\documentclass[%
 aip,
 amsmath,amssymb,
 reprint,%
]{revtex4-1}
\usepackage{graphicx}
\usepackage{dcolumn}
\usepackage{bm}

\usepackage[utf8]{inputenc}
\usepackage[T1]{fontenc}
\usepackage{mathptmx}
\usepackage{etoolbox}

\usepackage{vale}
\usepackage{mathtools}
\usepackage[varg]{txfonts}   
\usepackage{color}
\usepackage{colortbl}
\usepackage{array}
\usepackage{amssymb}
\usepackage{amsmath}
\usepackage{mathtools}
\usepackage{tabularx}
\usepackage{multirow}
\usepackage{subfigure}
\usepackage{psfrag}
\usepackage{paralist}
\usepackage{setspace}
\usepackage{comment}
\usepackage{morefloats}
\usepackage{vale}
\usepackage{natbib}
\usepackage{bm}
\usepackage[mathlines]{lineno}
\usepackage{listings}
\usepackage{xcolor}

\makeatletter
\def\@email#1#2{%
 \endgroup
 \patchcmd{\titleblock@produce}
  {\frontmatter@RRAPformat}
  {\frontmatter@RRAPformat{\produce@RRAP{*#1\href{mailto:#2}{#2}}}\frontmatter@RRAPformat}
  {}{}
}%
\makeatother
\begin{document}

\preprint{AIP/123-QED}

\title[]{A robust intermittency equation formulation for transition modeling in Spalart--Allmaras simulations of airfoil flows over a wide range of Reynolds numbers}

\author{Valerio D'Alessandro}
 \email{v.dalessandro@univpm.it.}
\affiliation{Dipartimento di Ingegneria Industriale e Scienze Matematiche, Università Politecnica delle Marche, Via Brecce Bianche 12, 60131 Ancona (AN), Italy}

\author{Matteo Falone}%
\affiliation{Dipartimento di Ingegneria Industriale e Scienze Matematiche, Università Politecnica delle Marche, Via Brecce Bianche 12, 60131 Ancona (AN), Italy}

\author{Luca Giammichele}%
\affiliation{Dipartimento di Ingegneria Industriale e Scienze Matematiche, Università Politecnica delle Marche, Via Brecce Bianche 12, 60131 Ancona (AN), Italy}

\author{Renato Ricci}%
\affiliation{Dipartimento di Ingegneria Industriale e Scienze Matematiche, Università Politecnica delle Marche, Via Brecce Bianche 12, 60131 Ancona (AN), Italy}

\date{\today}

\begin{abstract}
This paper introduces a new robust formulation for local correlation--based 
laminar--to--turbulent transition models.
This mechanism is incorporated into Reynolds--Averaged Navier--Stokes (RANS) equations, 
coupled with the Spalart–Allmaras (SA) turbulence model, considering 
both $\gamma$ and $\gamma$–$\Ret$ transition frameworks.
In this context, special attention is placed on numerical stabilization of the $\gamma$ transport equation,
which is identified as the root cause of instabilities observed in both
$\gamma$ and $\gamma$–$\Ret$ based models.
To this end, the intermittency equation is reformulated
in logarithmic form and further stabilized through an energy--based limiting to 
bound excessively high positive values.
In order to suppress unphysical pressure oscillations 
in the transition region, a gradient--driven artificial viscosity is also
introduced.
Additionally, the SA equation is augmented
with strain--rate modulated production and rotation correction terms.\\
The presented approach has demonstrated consistent effectiveness and 
robustness in the simulation of flow fields around airfoils 
over a wide range of Reynolds numbers, making it suitable for practical aerodynamic 
design applications.
%
\end{abstract}

\maketitle

\section{Introduction}\label{sec:intro}
\noindent The aim of this paper is to contribute to the ongoing research
 on the development of CFD techniques for predicting laminar--to--turbulent transition.\\
%
Specifically, this study focuses on flow over airfoils, where transition in the
 boundary layer, whether attached or separated, can affect significantly the 
overall flow behaviour.
%
In this regard, a reliable prediction of the flow's laminar--to--turbulent transition 
plays a crucial role, especially when Laminar Separation Bubbles (LSBs) 
occur on the airfoil surface~\cite{WINDTE:2006}. 
Moreover, accurate prediction of transitional phenomena is crucial for 
evaluating aerodynamic force coefficients, which are fundamental to the 
design process across a wide range of engineering disciplines.\\
In many cases, the flow field developing around airfoils 
is very complex, often requiring the adoption of turbulent scales resolving techniques 
such as Large--Eddy Simulation (LES), particularly at high angles of attack. 
LES currently remains impractical 
for industrial applications due to its high computational cost. 
Therefore, Reynolds--Averaged Navier--Stokes (RANS) equations based
approaches are the only feasible strategies that can be employed
into industrial workflows.
%
However, these methods are inadequate in transitional regimes, as they are typically developed under 
the assumption of a fully turbulent regime. To address this limitation, 
several locally formulated transition models have been integrated into the 
RANS framework over the years~\cite{MenterLangtry2:2006,MenterLangtry:2006,DKWalters:2006}.
A prominent example is the class of local correlation-based transition models (LCTMs),
which have demonstrated good accuracy across a broad range of applications.
LCTMs combine CFD--compatible transport equations with empirical correlations
for the purpose of transition modeling.\\
Notably, this paper aims to extend $\gamma$–$\Ret$ 
methodology~\cite{MenterLangtry2:2006,MenterLangtry:2006}, which is a pioneering LCTM approach,
and also to explore a simpler transition approach based on a single transport equation,~\emph{i.e.} the $\gamma$ model.\\ 
It is worth noting that the $\gamma$ and $\gamma$–$\Ret$ approaches were originally
 introduced for the $k$--$\omega$ turbulence model.
However, the present work investigates their implementation in combination with the
Spalart--Allmaras (SA) turbulence model~\cite{SApaper}.
This choice is particularly appealing, since it reduces
the number of partial differential equations (PDEs) that must be solved,
and thereby the computational resource requirements.
%
Consequently, this modelling technique attracted increasing research interest over the past decade.
Therefore, several versions of the LCTM, coupled with the 
SA turbulence model, have been 
developed~\cite{Medida:2011, Wang-SA, DAlessandro_sa, Nichols:2019}, 
based on two--equation transition models.
At the same time, some studies have also introduced  
single equation models for handling the laminar--to--turbulent transition~\cite{LIU2020106128, Bader-1eqn, Dalessandro:2021,Renac:2024}.\\
\textcolor{black}{It is important to note that only a few studies in the literature 
have explicitly addressed the numerical instabilities associated with LCTMs~\cite{Piotrowsky01,Piotrowsky02}.
This gap is noteworthy, as also confirmed by our computational experience, indicating that such instabilities 
are recurrent in LCTM frameworks due to the inherent numerical challenges of the intermittency transport equation.}
%
%
Accordingly, one of the main aims of this paper is to present 
and investigate a stabilized formulation for intermittency equation itself.
%
\textcolor{black}{
In detail, in our previous works~\cite{DAlessandro_sa,RIZZO2020105620,Dalessandro:2021} we encountered
 numerical instabilities associated with negative intermittency values.
For this reason, a positivity–preserving implementation based on the logarithmic 
formulation of the intermittency equation is adopted in the present study.
This approach has been only marginally explored in the literature for transition models: it was preliminarly 
applied by our research group for incompressible flows, while 
Plath et al.~\cite{Renac:2024} employed it for transitional transonic regimes.
Furthermore, an additional stabilization mechanism based on an energy--limiting procedure 
is introduced to constrain excessively high positive values in the logarithmic intermittency formulation.
This modification is required because the numerical solution process may generate unphysical overshoots
 beyond the physical range of intermittency, which can lead to numerical divergence.
Energy--limiting stabilization techniques were originally introduced in turbulence modeling by
 Allmaras et al.~\cite{Allmaras:2012} to prevent negative eddy--viscosity values in the SA equation.
In the present work, this concept is extended to the intermittency equation to control overly 
high positive values, thereby ensuring stable convergence.
}
In addition, 
\textcolor{black}
{
in our computational experiments we also observed Gibbs--like oscillations
in the pressure field within the transition region.
To mitigate this undesired behavior, a gradient--driven artificial
viscosity term was introduced into the intermittency equation,
This additional dissipation, which to the authors' knowledge has not been 
previously applied to the intermittency equation, effectively damps spurious pressure 
fluctuations without affecting the predicted transition behavior.
}
Notably, these approaches
are successfully applied to both  $\gamma$ and $\gamma$–$\Ret$ transition models.
It is important to emphasize that without the proposed stabilization and artificial viscosity, the computations does not 
converge or exhibit  non--physical pressure oscillations in the transition region.\\
The formulation of the SA equation was also addressed in this paper. 
More precisely, strain--modulated production~\cite{Rung:2003} and rotation-correction for 
the source terms~\cite{SA-rot} were carefully
assessed for their effectiveness in modelling separation--induced phenomena.\\
%
Finally, the reliability and robustness of the proposed approaches are demonstrated through the
 computation of several flow fields around different airfoils operating over a wide range of Reynolds numbers.\\
This paper is organized as follows: Sec.~\ref{sec:goveq} contains the governing
equations; the adopted numerical strategy and some relevant implementation details are presented in Sec.~\ref{sec:numerical}; 
the numerical results are discussed in Sec.~\ref{sec:res}; and Sec.~\ref{sec:concl} contains the conclusions.

\section{Governing equations} \label{sec:goveq}
In this section, we present the governing equations solved in this work.
All the flow models share the Spalart--Allmaras equation for turbulence modeling.
Transitional effects, on the other hand, are accounted for using
either the $\gamma$ and $\gamma$--$\Ret$ models.
%
\subsection{Turbulence model} \label{sec:turb}
\noindent The governing equations of incompressible RANS with the Spalart--Allmaras model are given by:
\begin{linenomath*}
\begin{equation} \label{eq:ns_v_cons}
  \begin{aligned}
&   \frac{\partial u_j}{\partial x_j}  = 0~, \\
&   \frac{\partial u_i}{\partial t}
  + \frac{\partial}{\partial x_j}\left(u_i u_j\right) = -\frac{\partial p}{\partial x_i} 
  + \frac{\partial}{\partial x_j} \left[ 
                                          2\left(\nu + \nu_t\right) S_{ij}  
                                  \right]~,\\
&   \frac{\partial \nutilde}{\partial t} 
  + \frac{\partial}{\partial x_j}\left(u_i \nutilde\right) =
    \mathrm{P}_{\nutilde} - \mathrm{D}_{\nutilde} 
   +\frac{c_{b2}}{\sigma} \frac{\partial \nutilde}{\partial x_j} \frac{\partial \nutilde}{\partial x_j}
   +\frac{\partial}{\partial x_j} \left[ 
                                          \left(\nu + \frac{\nutilde}{\sigma}\right)  \frac{\partial \nutilde}{\partial x_j} 
                                  \right]~, \\
  \end{aligned}
\end{equation}
\end{linenomath*}
\noindent where $u_i$ is the $i$--th component of the velocity vector, $p = P/\rho$ is the pressure 
divided by the density, $S_{ij}$ denotes the strain rate tensor, and $\nu$ is the kinematic viscosity.
The production and destruction terms for $\nutilde$ are evaluated as follows:
\begin{linenomath*}
\begin{equation} \label{eq:nutilde-source}
  \begin{aligned}
   &\mathrm{P}_{\nutilde} = c_{b1}\tilde{S}\nutilde~, \\
   &\mathrm{D}_{\nutilde} = c_{w1}f_w \frac{\nutilde^2}{d^2}~,
  \end{aligned}
\end{equation}
\end{linenomath*}
\noindent in which $d$ denotes the distance to the nearest wall. Conversely, 
the turbulent viscosity, $\nu_{t}$, is computed based on the $\tilde{\nu}$ variable as follows:
\begin{linenomath*}
\begin{equation}
\nu_t= f_{v1} \nutilde~.
\end{equation}
\end{linenomath*}
%
%
\noindent For $\mathrm{P}_{\nutilde}$ and $\mathrm{D}_{\nutilde}$ terms (eq.~\ref{eq:nutilde-source}), the following closure functions are used: 
%
\begin{linenomath*}
\begin{equation} \label{eq:clfun2}
\begin{aligned}
g&= r + c_{w2}\left( r^6 -r \right), \quad f_{w}= g\left[ \frac{1+c_{w3}^6}{g^6 + c_{w3}^6} \right]^{\frac{1}{6}}, \\
& f_{v1} = \frac{\chi^3}{\chi^3 + c_{v1}^3}~, \quad
  f_{v2} = 1 - \frac{\chi}{1 + \chi f_{v1}}~, \\
& \Stilde = \left[\Omega + \min\left(0, S - \Omega\right)\right] + \frac{\nutilde}{k^2 d^2} f_{v2}~, \\
& r = 
\begin{cases}
-\frac{\nutilde}{\Stilde k^2 d^2}, & \text{if } \frac{\nutilde}{\Stilde k^2 d^2} < 0 \\[6pt]
\min\left(\frac{\nutilde}{\Stilde k^2 d^2}, r_{max}\right)~, & \text{if } \frac{\nutilde}{\Stilde k^2 d^2} \ge 0
\end{cases} ~.
\end{aligned}
\end{equation}
\end{linenomath*}
\noindent Here, $\chi = \tilde{\nu} / \nu$ is the dimensionless turbulent 
variable, $\Omega$ is the magnitude of the vorticity tensor and $S$ is 
the magnitude of the strain rate tensor.\\
To fully define the turbulence model, the following standard closure constants are employed:
\begin{linenomath*}
\begin{alignat}{3} \label{eq:clcos}
 &c_{b1}=0.1355~,& \quad &c_{b2}=0.622~,& \quad &c_{v1}=7.1~, \\
 &c_{w1}=\frac{c_{b1}}{k^2} + \frac{(1+c_{b2})}{\sigma}~,&  &c_{w2}=0.3~,& &c_{w3}=2~,  \\
 &\sigma=2/3~,& \quad &r_{max}=10~,& \quad &k=0.41~. 
\end{alignat}
\end{linenomath*}
%
%

\subsubsection{Spalart--Allmaras model versions}
The SA model described above is hereinafter referred to as the SA model with rotation correction, 
and it will be denoted as the SA--R1 model version. 
To be specific, the expression for $\Stilde$ given in eq.~\ref{eq:clfun2} is derived from De Dacles-Mariani et al.~\cite{SA-rot}.
This formulation attempts to empirically adjust the production term for vortex dominated flow reducing the eddy--viscosity 
in the regions where the vorticity is grater than the strain rate.
This is particularly relevant in the context of the present paper, 
as the suppression and enhancement of turbulence production can lead to vortex breakdown
 especially when the vortex is shed from a separated shear layer. 
These flow features are highly recurrent on airfoils, making this approach particularly appealing.
The model version is referred to as R1 since the coefficient before the minimum condition in $\Stilde$
(eq.~\ref{eq:clfun2}) is equal to one (note that in the literature this value is not strictly fixed).\\
Furthermore, this paper employs a second variant of the SA model, known as the strain--modulated 
production approach (referred to as SA–SMP).
In this formulation, the quantity $\Stilde$ in the production term is 
replaced with the following expression:
\begin{linenomath*} 
\begin{equation}  \label{eq:clfun3}
\Stilde = S \left( \frac{1}{\chi}  + f_{v1}  \right)~,
\end{equation}
\end{linenomath*}
a modification inspired by the work of Baldwin and Barth~\cite{baldwin1990one}.
It is important to note that $r$ function is still evaluated, in the SA--SMP version, using 
the $\Stilde$ expression from SA--R1 formulation.
%
Notably, SA--R1 and SA--SMP utilize all the
equations and constant described in Sec.~\ref{sec:turb}
with the sole exception of $\Stilde$ function.
\textcolor{black}{
The idea behind this approach
is to ensure the physical modeling in the wall region through the standard $r$ function, while
the production term benefits from improved stability and accuracy in vortex-dominated regions,
particularly relevant for transitional shear layers over airfoils.
This design follows the original rationale of Baldwin and Barth, who first proposed the strain–modulated
production to improve robustness of eddy–viscosity closures without requiring recalibration of model constants.
}
\subsection {Local correlation--based transition models}
\subsubsection {Two--equation model}
\noindent As a first strategy, two additional transport equations 
are employed to model the transition, namely $\gamma$ and $\Ret$, 
which represent the intermittency and the transition onset momentum thickness 
Reynolds number, respectively. 
\textcolor{black}{The present transition model builds upon the formulation proposed in our previous work~\cite{DAlessandro_sa}, 
with two main modifications: a revised expression for the intermittency production term and 
an updated coupling strategy between the SA turbulence equation and the transition model, 
as detailed later in this subsection. 
Accordingly,} the equations are given by:
\begin{linenomath*}
\begin{equation} \label{eq:gamma_re}
  \begin{aligned}
&\frac{\partial \gamma}{\partial t} +  \frac{\partial}{\partial x_j} \left(u_j \gamma\right)   =
 \mathrm{P}_\gamma  - \mathrm{D}_\gamma 
+ \frac{\partial}{\partial x_j}\left[\left( \nu + {\nu_t}  \right) \frac{\partial \gamma  }{\partial x_j}  \right]~, \\
&\frac{\partial \Ret}{\partial t} + \frac{\partial}{\partial x_j} \left(u_j \Ret\right) =
\mathrm{P}_{\tetat} + \frac{\partial}{\partial x_j}\left[\sigma_{\tetat}  \left(  \nu + {\nu_t} \right) \frac{\partial \Ret  }{\partial x_j}  \right]~.
  \end{aligned}
\end{equation}
\end{linenomath*}
The source terms in the $\gamma$ equation take the expressions:
\begin{linenomath*}
\begin{equation} \label{eq:gamma_source}
  \begin{aligned}
&\mathrm{P}_\gamma = c_{a1}~S~\mathrm{F_{length}} \left[ \mathrm{F_{onset}} \right]^{0.5} \gamma \left(1 - c_{e1} \gamma \right)~, \\
&\mathrm{D}_\gamma = c_{a2}~\Omega~\mathrm{F_{turb}}~\gamma \left(c_{e2} \gamma -1\right)~, 
  \end{aligned}
\end{equation}
\end{linenomath*}
in $\mathrm{P}_\gamma$ the term $\mathrm{F_{onset}}$ is computed as:
\begin{linenomath*}
\begin{equation} \label{eq:fonset}
\mathrm{F_{onset}} =  \mathrm{max} \left( \mathrm{F_{onset,2}} - \mathrm{F_{onset,3}}, 0  \right)  
\end{equation}
\end{linenomath*}
with 
\begin{linenomath*}
\begin{equation} \label{eq:fonset1}
  \begin{aligned}
& \mathrm{F_{onset,2}} =  \mathrm{min} \left( \mathrm{max} \left( \mathrm{F_{onset,1}} , \mathrm{F_{onset,1}}^4  \right)  , 4\right)~, \\
& \mathrm{F_{onset,3}} =  \mathrm{max} \left( 2 - \left( \frac{\mathrm{R_T}}{2.5}  \right)^3 ,0 \right)~, \\
& \mathrm{F_{onset,1}} = \frac{\mathrm{Re}_\nu}{2.193 \mathrm{Re}_{\theta,c}}~.\\
  \end{aligned}
\end{equation}
\end{linenomath*}
The terms $\mathrm{Re}_\nu$ and $\mathrm{R_T}$ in eq.~\ref{eq:fonset1} are determined as follows:
\begin{linenomath*}
\begin{equation}
  \mathrm{Re}_\nu  = \frac{S d^2}{\nu} \qquad  \mathrm{R_T} = \frac{\nu_t}{\nu}~.
\end{equation}
\end{linenomath*}
%
%
%
Regarding $\mathrm{D}_\gamma$, the coefficient $\mathrm{F_{turb}}$ 
is given by:
\begin{linenomath*}
\begin{equation}
  \mathrm{F_{turb}} = \exp\left(-\frac{\mathrm{R_T}}{4}\right)^4 .
\end{equation}
\end{linenomath*}
%
The source terms in the transport equation for $\Ret$ are modeled using the following expression for $\mathrm{P}_{\tilde{\theta_t}}$:
\begin{linenomath*}
\begin{equation}\label{eq:ptetat}
  \mathrm{P}_{\tetat}  = \frac{c_{\theta,t}}{\mathrm{T}} \left( \mathrm{Re}_{\theta,t} - \Ret \right) \left(1- \mathrm{F}_{\theta,t} \right)~.
\end{equation}
\end{linenomath*}
In eq.~\ref{eq:ptetat}, the final term, $\mathrm{F}_{\theta,t}$, is formulated as:
\begin{linenomath*}
\begin{equation}\label{eq:ptetat2}
  \mathrm{F}_{\theta,t} = \mathrm{min} \left( 
                                              \mathrm{max} \left( \exp\left(-\frac{ \left|\uvec\right|^2   }{375\Omega \nu \Ret}\right)^4 ,
                                                                  1 - \left( \frac{\gamma -1/c_{e2}}{1-1/c_{e2}} \right)^2
                                                           \right)
                                              ,
                                              1.0
                                       \right)~. 
\end{equation}
\end{linenomath*}
%
%
The term $\mathrm{T}$ appearing in $\mathrm{P}_{\tilde{\theta_t}}$ 
can be expressed as: $500\nu / \left|\uvec\right|^2$.
The treatment of $\mathrm{Re}_{\theta,t}$ in eq.~\ref{eq:ptetat}, along with the coefficient $\mathrm{F_{length}}$, 
is discussed in the following subsection.\\
It is also important to note that for the transition model, the following 
closure constants are adopted:
\begin{linenomath*}
\begin{alignat}{4} \label{eq:clcos}
 c_{a1}&=2.0~,& \quad c_{a2}&=0.06~,&        \quad c_{e1}&=1.0~, \\
 c_{e2}&=50~,& \quad c_{\theta,t}&=0.03~,&  \quad \sigma_{\theta,t}&=2.0~.& 
\end{alignat}
\end{linenomath*}
%
%
%
\noindent Like other $\gamma$--$\Ret$ transition models, 
the present model includes three empirical correlations required to compute 
$\mathrm{Re}_{\theta,t}$, $\mathrm{Re}_{\theta,c}$, and $\mathrm{F}_{\text{length}}$.
Detailed explanations of these terms can be found in literature~\cite{BOUCHARD2021107045}
and are not discussed further, as this lies outside the scope of the present paper.\\
%
%
%
%
%
%
In this work, the correlation for $\mathrm{Re}_{\theta,t}$ is taken directly from 
the work of Menter et al.~\cite{MenterLangtry2:2006} :
\begin{linenomath*}
\begin{equation} \label{eq:rethetat_corr_x}
\begin{aligned}
&\mathrm{Re}_{\theta,t1} = 1173.51  - 589.428 \cdot Tu + 0.2196/Tu^2~,  \\ 
&\mathrm{Re}_{\theta,t2} = 331.5\left(Tu - 0.5668\right)^{-0.671}~, \\ 
&\mathrm{Re}_{\theta,t} = \left\{ \begin{array}{l}
\mathrm{Re}_{\theta,t1}~F\left(\lambda_\theta \right) \quad Tu \leq 1.3\\
\mathrm{Re}_{\theta,t2}~F\left(\lambda_\theta \right) \qquad  Tu>1.3
\end{array} \right. , 
\end{aligned}
\end{equation}
\end{linenomath*}
\begin{linenomath*}
\begin{equation}\label{eq:rethetat_corr_y}
\begin{aligned}
&F_1 = 12.986 \lambda_\theta + 123.66 \lambda_\theta^2 + 405.689 \lambda_\theta^3~, \\
&F_2 = 0.275\left[1-\exp\left( -35 \lambda_\theta \right)\right]~, \\
&F\left(\lambda_\theta \right) = \left\{ \begin{array}{l}
1 + F_1 \exp\left( - \left( \frac{Tu}{1.5}  \right)^{1.5}\right)  \quad \lambda_\theta \leq 0\\
1 + F_2 \exp\left( -\frac{Tu}{0.5}  \right)   \qquad  \lambda_\theta >0
\end{array} \right. . 
\end{aligned}
\end{equation}
\end{linenomath*}
%
%
\noindent It is important to note that the correlations 
in eq.~\ref{eq:rethetat_corr_x} and~\ref{eq:rethetat_corr_y} 
involve the turbulence intensity, $\mathrm{Tu}$.
Within the framework of the $k$–$\omega$ model, $\mathrm{Tu}$ can
 be evaluated using the solution of the $k$-equation.
This work is based on the SA equation for turbulence modeling. 
Consequently, establishing a direct relationship between the 
turbulence intensity $\mathrm{Tu}$ and the working variable $\nutilde$ 
is not straightforward. 
For this reason,  we adopt the approach proposed by Medida and Baeder~\cite{Medida:2011}
in which the turbulence intensity is fixed throughout the flow field. 
Specifically, we set $\mathrm{Tu} = \mathrm{Tu}_\infty$ at all computational cells.\\
Moreover, $\mathrm{Re}_{\theta,t}$ is computed by iterating 
on the value of $\theta_t$, since it depends on $\theta_t$
itself due to the presence of $\lambda_{\theta}$.
By contrast, for $\mathrm{Re}_{\theta,c}$ and $\mathrm{F}_{\text{length}}$,
 we adopt the correlations developed by Malan et al.~\cite{AIAA-Malan}:
%
%
%
\begin{linenomath*}
\begin{equation}
\mathrm{Re}_{\theta,c} = \mathrm{min} \left( 0.615\Ret + 61.5 , \Ret \right)~, 
\end{equation}
\end{linenomath*}
\begin{linenomath*}
\begin{equation}
\mathrm{F_{length}} = \mathrm{min} \left( \exp\left( 7.168-0.01173\Ret  \right) + 0.5 , 300\right)~.
\end{equation}
\end{linenomath*}
\\
%
%
\\
\noindent The production and destruction terms appearing in the $\nutilde$,
and outlined in eq.~\ref{eq:nutilde-source}, are suitably modified as follows
in order to couple the turbulence and transition models:
\begin{linenomath*}
\begin{equation} \label{eq:nutilda2_src}
     \begin{aligned}
      &\mathrm{P}_{\nutilde} = \gamma_{\mathrm{eff}}  c_{b1}\Stilde \nutilde~, \\
      &\mathrm{D}_{\nutilde} = 
                               \min \left(
                                            \max\left(\gamma , \textcolor{black}{\beta}\right) , 1.0 
                                    \right)
c_{w1} f_{w} \left( \frac{\nutilde}{{d}} \right)^2~,
     \end{aligned}
\end{equation}
\end{linenomath*}
\textcolor{black}{in the above equation, $\beta$ is set to 0.1 instead of 0.5 as adopted in our previous work~\cite{DAlessandro_sa}.}
The term $\gamma_{\mathrm{eff}}$ in eq.~\ref{eq:nutilda2_src}, 
which accounts for separation--induced transition, is defined as follows:
\begin{linenomath*}
\begin{equation} \label{eq:gamma_sep}
\gamma_{\mathrm{eff}} = \mathrm{max} \left( \gamma, \gamma_{\mathrm{sep}} \right)
\end{equation}
\end{linenomath*}
with
\begin{linenomath*}
\begin{equation} \label{eq:gamma_sep}
  \begin{aligned}
        \gamma_{\mathrm{sep}} = \mathrm{min} \left( 2.0 ~ 
                                                      \mathrm{max} \left[ 0 , \left( \frac{\mathrm{Re}_\nu}{3.235\mathrm{Re}_{\theta,c}} \right) -1\right]\mathrm{F}_{reattach}, 2.0
                                             \right)\mathrm{F}_{\theta,t}
  \end{aligned}
\end{equation}
\end{linenomath*}
and
\begin{linenomath*}
\begin{equation} \label{eq:f_reatt}
\mathrm{F}_{reattach} = \exp\left(-\frac{\mathrm{R_T}}{20}\right)^4.
\end{equation}
\end{linenomath*}
Furthermore, intermittency transport equation is re--written,
in logarithmic form, where $\tilde{\gamma} = \log \gamma$.
This strategy, originally proposed by Ilinca and Pelletier~\cite{ILINCA1999560} for turbulence modeling, is particularly appealing because it inherently ensures
the positivity of the variable.
In addition, the smoother distribution of the logarithmic variable, compared to the original, contributes to enhanced numerical robustness
and stability.
The equation for $\tilde{\gamma}$ is derived by substituting $\gamma = \exp(\tilde{\gamma})$ into the original intermittency equation.
The resulting transport equation for $\tilde{\gamma}$ is:
\begin{equation} \label{eq:gamma_re-log}
\frac{\partial \gammatilde}{\partial t} +  \frac{\partial}{\partial x_j} \left(u_j \gammatilde\right)   =
 \mathrm{P}_{\gammatilde}  - \mathrm{D}_{\gammatilde} 
+\frac{\partial}{\partial x_j}\left[\left( \nu + {\nu_t}  \right) \frac{\partial \gammatilde  }{\partial x_j}  \right] 
+\left( \nu + {\nu_t}  \right) \frac{\partial \gammatilde  }{\partial x_j} \frac{\partial \gammatilde  }{\partial x_j} \\
\end{equation}
with
\begin{linenomath*}
\begin{equation} \label{eq:gamma_source-log}
  \begin{aligned}
&\mathrm{P}_{\gammatilde} = c_{a1}~S~\mathrm{F_{length}} \left[ \mathrm{F_{onset}} \right]^{0.5}  \left(1 - c_{e1} e^{\gammatilde} \right) ~, \\
&\mathrm{D}_{\gammatilde} = c_{a2}~\Omega~\mathrm{F_{turb}} \left(c_{e2} e^{\gammatilde} -1\right)~, 
  \end{aligned}
\end{equation}
\end{linenomath*}
this is a key aspect of the approach presented in this paper,
as it helps prevent the blow--up of the computations we have experienced
with the baseline formulation.\\
Note also that in the standard formulation of the $\gamma$--$\Ret$ model, the production
term of the intermittency includes a square--root dependency by itself,
namely $\mathrm{P}_{\gamma} \propto \gamma^{0.5} \left(1 - c_{e1}\gamma\right)$.
Based on our computational experience, we have observed that this term poses significant
challenges when the equation is formulated in logarithmic space.
In particular, $\gamma^{0.5}$ transforms into an exponential factor, $e^{-0.5 \gammatilde}$,
which leads a rapid increase of the production term in the zones where $\gammatilde <0$. 
This behaviour severely compromise the numerical stability of PDE 
and often causes divergence.
To overcome this issue, the production term in eq.~\ref{eq:gamma_source} was reformulated
by removing $\gamma^{0.5}$ dependency, yielding a linear source term, $\mathrm{P}_{\gamma} \propto \gamma \left(1 - c_{e1}\gamma\right)$,
which leads to the term source term shown in eq.~\ref{eq:gamma_source-log}.
This modification is essential for achieving convergence
across the range of the considered flow cases.
\textcolor{black}{
Importantly, this adjustment does not alter the physical closure of the model, as also confirmed by the computational results presented in Sec.~\ref{sec:res}.
A similar modification has been previously adopted in the literature~\cite{Nichols:2019,LIU2020106128, Dalessandro:2021, Renac:2024}, although without providing a detailed rationale.
}
\subsubsection{One--equation model}
Additionally, a second strategy for including
transitional effects in Spalart--Allmaras 
turbulence model is investigated in this work. 
Particularly, only a single transport equation for $\gammatilde = \log \gamma$,
identical to that presented in eq.~\ref{eq:gamma_re-log}, is solved.
In contrast, the transition onset location is determined through algebraic correlations.
\textcolor{black}{
In the present study, these correlations follow the formulation proposed by Liu et al.~\cite{LIU2020106128}, 
with a minor modification described in the following.
}
$\mathrm{F_{length}}$ is assigned to a fixed value of $0.5$. At the same time,
some modifications, derived from the approach proposed by Liu et al.~\cite{LIU2020106128}, are introduced to the empirical 
terms present in the source terms  of the intermittency equation. The expression for $\mathrm{F_{onset}}$ is
is inherited from $\gamma$--$\Ret$ model, while $\mathrm{F_{onset,2}}$ and $\mathrm{F_{onset,3}}$
 are re--calculated as follows:
\begin{linenomath*}
\begin{equation} \label{eq:fonset1_log}
  \begin{aligned}
& \mathrm{F_{onset,2}} =  \mathrm{min} \left(  \mathrm{F_{onset,1}} , 2.0\right)~, \\
& \mathrm{F_{onset,3}} =  \mathrm{max} \left( 1 - \left( \frac{\mathrm{R_T}}{3.5}  \right)^3 ,0 \right)~. \\
  \end{aligned}
\end{equation}
\end{linenomath*}
%
A relevant difference between one--equation and two--equation models concerns $\mathrm{Re}_{\theta,c}$ which
is now computed using the following correlations:
\begin{linenomath*}
\begin{equation} \label{eq:retc1}
  \mathrm{Re}_{\theta,c} = f_{local} \left( Tu_l \right)  f_{far} \left( Tu_\infty \right)~,
\end{equation}
\end{linenomath*}
where 
\begin{linenomath*}
\begin{equation} \label{eq:retc2} 
 \begin{aligned} 
   &  f_{local} \left( Tu_l \right) = 803.73 \left( Tu_l + 0.6067 \right)^{-1.027}~, \\ 
   &  f_{far} \left( Tu_\infty \right)  = -3.162 Tu_\infty^2 - 0.4565 Tu_\infty + 1.7~.
 \end{aligned}
\end{equation}
\end{linenomath*}
A particular feature of eq.~\ref{eq:retc2} is the inclusion, $Tu_l$,
through the correlated term $f_{local}$. This term is treated 
following Liu et al.~\cite{LIU2020106128}, 
using the formulation discussed by Cakmakcioglu~\cite{local_tu} for 
the computation of $Tu_l$.\\
As regards $\mathrm{F_{turb}}$, which is related to intermittency destruction term, 
the following relation is used:
\begin{linenomath*}
\begin{equation}
  \mathrm{F_{turb}} = \exp\left(-\frac{\mathrm{R_T}}{2}\right)^4~.
\end{equation}
\end{linenomath*}
%
Similarly to the two--equation model $\mathrm{P}_{\nutilde}$ and $\mathrm{D}_{\nutilde}$
are appropriately modified to embed transitional effects into the turbulence model:
\begin{linenomath*}
\begin{equation} \label{eq:nutilda2_src2}
     \begin{aligned}
      &\mathrm{P}_{\nutilde} = \gamma_{\mathrm{eff}}  c_{b1}\Stilde \nutilde~, \\
      &\mathrm{D}_{\nutilde} = 
                                            \max\left(\gamma , 0.1\right) 
                                            c_{w1} f_{w} \left( \frac{\nutilde}{{d}} \right)^2~,
     \end{aligned}
\end{equation}
\end{linenomath*}
where
\begin{linenomath*}
\begin{equation}\label{eq:log_gamma2}
\gamma_{sep} = \min \left(  8.0 \max \left[  0.0 , \left(  \frac{\mathrm{Re}_\nu}{2 \mathrm{Re}_{\theta,c}}  \right) -1  \right]    , \gamma_{lim}\right)
\end{equation}
\end{linenomath*}
this follows the correlation suggested by Liu et al.~\cite{LIU2020106128}, except that here $\gamma_{lim} = 2.5$ 
instead of $1.0$ used in the seminal contribution.
\textcolor{black}{Finally, it should be noted that the coupling between the intermittency and $\tilde{\nu}$ appearing in the destruction term of eq.~\ref{eq:nutilda2_src2} is inspired by the approach of Plath et al.~\cite{Renac:2024}.}
\\
%
\subsection{Stabilization of intermittency equation}
\subsubsection{Energy limiting }
\noindent The logarithmic formulation for intermittency prevents
instabilities caused by sign changes in the variable; however, by itself, it is    
not sufficient to fully stabilize the numerical solution.
%
In fact, a closer examination
 of the equation reveals that it lacks any mechanism to prevent
 excessively high positive values of $\tilde{\gamma}$, which can trigger floating--point exceptions.
In this context, we have devised a method to bound $\tilde{\gamma}$, inspired by the
work of Allmaras et al.~\cite{Allmaras:2012} which was intended to prevent negative values of $\nutilde$.
%
%
Particularly, the $\tilde{\gamma}$ transport equation is multiplied on 
both sides by the variable itself, yielding:
\begin{equation} \label{eq:gamma_re-log2}
\begin{aligned}
\frac{\partial }{\partial t} \left(\frac{\gammatilde^2}{2}\right) 
&+  \frac{\partial}{\partial x_j} \left(u_j  \frac{\gammatilde^2}{2}  \right)  =\\
&=\gammatilde \left( \mathrm{P}_{\gammatilde}  - \mathrm{D}_{\gammatilde}  \right)
+\gammatilde \frac{\partial}{\partial x_j}\left[\left( \nu + {\nu_t}  \right) \frac{\partial \gammatilde  }{\partial x_j}  \right] 
+\gammatilde\left( \nu + {\nu_t}  \right) \frac{\partial \gammatilde  }{\partial x_j} \frac{\partial \gammatilde  }{\partial x_j} \\
\end{aligned}
\end{equation}
the above equation can be re--written as:
\begin{equation} \label{eq:gamma_re-log3}
\begin{aligned}
&\frac{\partial \tilde{e}_\gamma}{\partial t}  
+  \frac{\partial}{\partial x_j} \left(u_j  \tilde{e}_\gamma   \right)  =\\
&=\gammatilde \left( \mathrm{P}_{\gammatilde}  - \mathrm{D}_{\gammatilde}  \right)
+ \frac{\partial}{\partial x_j}\left[\left( \nu + {\nu_t}  \right) \frac{\partial \tilde{e}_\gamma  }{\partial x_j}  \right] 
+\left(\gammatilde -1 \right)\left( \nu + {\nu_t}  \right) \frac{\partial \gammatilde  }{\partial x_j} \frac{\partial \gammatilde  }{\partial x_j} 
\end{aligned}
\end{equation}
where $\tilde{e}_\gamma$ denotes the artificial energy associated with $\gammatilde$,
defined as $\tilde{e}_\gamma = \gammatilde^2/2$. \\
We now define $\Omega^+$ and $\Omega^-$ as subdomains 
of the original domain $\Omega$
where $\tilde{\gamma}$ is positive 
and negative, respectively. 
Considering the case where $\partial \Omega^+ \cap \partial \Omega = \emptyset$,
and assuming that $\gammatilde$ is continuous,
it follows that $\left. \gammatilde\right|_{\partial \Omega^+} = 0$. A similar conclusion
applies for $\tilde{e}_\gamma$.\\
%
%
Therefore, the integration of eq.~\ref{eq:gamma_re-log3} over the 
subdomain domain $\Omega^+$ leads to the following expression:
%
\begin{equation} \label{eq:gamma_re-log4}
\begin{aligned}
&\int_{\Omega^+} \frac{\partial \tilde{e}_\gamma}{\partial t} d {\Omega} +
\int_{\partial\Omega^+}  u_j  \tilde{e}_\gamma d {S_j} = \\ 
&\int_{\Omega^+} \gammatilde \left( \mathrm{P}_{\gammatilde}  - \mathrm{D}_{\gammatilde}  \right) d {\Omega} +  
\int_{\partial\Omega^+} \left( \nu + {\nu_t}  \right) \frac{\partial \tilde{e}_\gamma  }{\partial x_j}   d {S_j} + \\ 
+ &\int_{\Omega^+} \left(\gammatilde -1 \right)\left( \nu + {\nu_t}  \right) \frac{\partial \gammatilde  }{\partial x_j} \frac{\partial \gammatilde  }{\partial x_j} d {\Omega}
\end{aligned}
\end{equation}
where the Gauss--Green divergence theorem is applied to simplify the equation. 
Since, by continuity, $\tilde{e}_\gamma$  and its gradient vanish on the boundary of $\Omega^+$, 
it follows that:
\begin{equation} \label{eq:gamma_re-log4}
\begin{aligned}
\int_{\Omega^+} \frac{\partial \tilde{e}_\gamma}{\partial t} d {\Omega} 
=&\int_{\Omega^+} \gammatilde \left( \mathrm{P}_{\gammatilde}  - \mathrm{D}_{\gammatilde}  \right) d {\Omega}~+\\
+ &\int_{\Omega^+} \left(\gammatilde -1 \right)\left( \nu + {\nu_t}  \right) \frac{\partial \gammatilde  }{\partial x_j} \frac{\partial \gammatilde  }{\partial x_j} d {\Omega}~.
\end{aligned}
\end{equation}
Introducing the variable $\tilde{E}_\gamma$ which is the overall artificial energy associated with $\gammatilde$
on $\Omega^+$:
\begin{equation} \label{eq:gamma_re-log5}
\tilde{E}_\gamma = \int_{\Omega^+} \tilde{e}_\gamma  d {\Omega}
\end{equation}
to avoid unbounded positive values of $\gammatilde$ (in terms of magnitude) within
 ${\Omega^+}$ the following constraint is applied:
\begin{equation} \label{eq:gamma_re-log6}
\frac{d \tilde{E}_\gamma}{d t} < 0~.
\end{equation}
Taking into account eq.~\ref{eq:gamma_re-log6} and 
eq.~\ref{eq:gamma_re-log4} it is possible to derive the following 
constraints:
\begin{equation} \label{eq:gamma_re-log7}
\begin{aligned}
&\gammatilde \left( \mathrm{P}_{\gammatilde}  - \mathrm{D}_{\gammatilde}  \right) < 0 \\
&\left(\gammatilde -1 \right)\left( \nu + {\nu_t}  \right) \frac{\partial \gammatilde  }{\partial x_j} \frac{\partial \gammatilde  }{\partial x_j} < 0
\end{aligned}
\end{equation}
the first condition is intrinsically satisfied since $\gammatilde > 0$, which implies, $\mathrm{P}_{\gammatilde}<\mathrm{D}_{\gammatilde}$.
By inspecting eq.~\ref{eq:clcos} and~\ref{eq:gamma_source-log}, it becomes clear that this condition is already fulfilled.
Conversely, the second requirement in eq.~\ref{eq:gamma_re-log7} leads to:
\begin{equation} \label{eq:gamma_re-log8}
\gammatilde < 1~,
\end{equation}
this condition ensures that ${d \tilde{E}_\gamma}/{d t} < 0$, since in our 
implementation the turbulent viscosity is constrained to non--negative values.
Note that $\gammatilde$ bounding is enforced through a simple clipping procedure
 applied as the final stage of the solution algorithm.
If this clipping is removed, stable computations become very difficult to achieve.\\
Finally, it is important to highlight that the bounding procedure just described
also applies to steady-state flows. In fact, if the same approach is adopted under steady
conditions, the left--hand side of eq.~\ref{eq:gamma_re-log4} represents the residual of $\gammatilde$,
~\emph{i.e.} its imbalance relative to a steady condition during the iterative procedure.
Hence, eq.~\ref{eq:gamma_re-log8} can be interpreted as a condition to be used to prevent 
nonphysical behavior during solver iterations.
\subsubsection{Gradient--driven artificial viscosity}
The PDE systems described in this section are clearly significantly stiff 
and their numerical solution is not a trivial task and the above
discussed stabilization strategy is crucial to obtain a solution.
However, this is not a complete remedy for the numerical instabilities
observed in $\gammatilde$ equation. Indeed, in several cases and as will be shown
below, a Gibbs--like phenomenon has been observed near 
the knee of the pressure distribution along the airfoil chord.
This behavior is clearly nonphysical. To mitigate it, we apply a gradient--driven artificial 
viscosity term in the intermittency equation, aimed at suppressing spurious pressure fluctuations.
Therefore, eq.~\ref{eq:gamma_re-log} is re--written as follows:
\begin{equation} \label{eq:gamma_re-log2b}
\begin{aligned}
\frac{\partial \gammatilde}{\partial t} +  \frac{\partial}{\partial x_j} \left(u_j \gammatilde\right)   =
 \mathrm{P}_{\gammatilde}  - \mathrm{D}_{\gammatilde} 
&+\frac{\partial}{\partial x_j}\left[\left( \nu + {\nu_t} +\nu_{art}   \right) \frac{\partial \gammatilde  }{\partial x_j}  \right] +  \\
&+\left( \nu + {\nu_t}  \right) \frac{\partial \gammatilde  }{\partial x_j} \frac{\partial \gammatilde  }{\partial x_j} \\
\end{aligned}
\end{equation}
where
\begin{equation} \label{eq:v_art}
\nu_{art} = c_{art}~\Delta \left| \NABLA \gammatilde \right|~.
\end{equation}
In eq.~\ref{eq:v_art} is a $c_{art}$ is the artificial viscosity scaling coefficient, 
while $\Delta$ is the characteristic length of the grid's cell. 
In this work, $\Delta$ is evaluated as the cube--root of the
 cell volume,~\emph{i.e.} $\Delta = \sqrt[3]{V_p}$ and 
we employ a fixed values of $c_{art}$ chosen empirically
 for each turbulence model variant. \\
Another relevant aspect of artificial viscosity concerns its impact on the energy limiting technique 
adopted to stabilize the intermittency transport equation.
Indeed, considering eq.~\ref{eq:gamma_re-log2b}, the second constraint of eq.~\ref{eq:gamma_re-log7}
modifies as follows:
\begin{equation} \label{eq:gamma_re-log7b}
\left[\left(\gammatilde -1 \right)\left( \nu + {\nu_t}  \right) - \nu_{art} \right] \frac{\partial \gammatilde  }{\partial x_j} \frac{\partial \gammatilde  }{\partial x_j} < 0~.
\end{equation}
In the above equation, it is straightforward to observe that $\nu_{art}$ acts a penalization term, thus the condition $\gammatilde < 1$
not only remains valid but is actually reinforced.
%
\subsection{Boundary conditions}
The boundary conditions for $\tilde{\nu}$ follow standard conventions:
$\nutilde_\infty = 3\nu$ is used in the free--stream, while $\tilde{\nu} = 0$ is enforced at the wall.
For the intermittency, a zero normal gradient condition is applied at the wall, and a fixed value of
$\gamma = 1$ (\emph{i.e.}, $\gammatilde  = 0$) is prescribed at the inlet.
Regarding $\Ret$, a zero--flux condition is applied at the wall, while a fixed--value condition is specified at the inlet. 
This inlet value is typically determined based on the local turbulence intensity.\\
\noindent All computational grids used in the following computations
ensure a viscous sub--layer scaled first cell height 
$y^{+} \approx 1$, as suggested in literature~\cite{MenterLangtry:2006}.
The value of $y^{+}$ is estimated as
$y^{+}= \frac{u_\tau}{\nu} y_c$,
where $u_\tau=\sqrt{\tau_w/\rho}$ is the friction velocity, $\tau_w$ is the viscous stress component
measured at the wall, and $y_c$ is the height of the cells next to the wall.\\

\section{Numerical approximation}~\label{sec:numerical}
The governing equations described in Sec.\ref{sec:goveq} 
are space discretized using an unstructured, collocated finite volume method (FVM). 
In particular, our solution strategy relies on the \texttt{foam--extend} v.5.0 library, 
a fork of the well--known OpenFOAM code, which is strongly focused on integrating 
community contributions.
It is worth noting that numerical solutions are obtained using \texttt{simpleFoam}, 
the steady--state solver for incompressible flows available in the official OpenFOAM releases. 
\texttt{simpleFoam} employs the well--established SIMPLE algorithm~\cite{Patankar:1980} 
for pressure--velocity coupling. To suppress pressure--velocity decoupling and eliminate spurious oscillations, 
the Rhie--Chow interpolation method is applied~\cite{Ferziger}.\\
For all computations presented in this work, the diffusive terms and 
pressure gradients were approximated using second--order accurate central scheme.
Our baseline strategy for discretizing the convective terms in the momentum, turbulence, 
and transition equations employs a second-order accurate linear--upwind scheme.
Regarding the linear solvers, a Preconditioned Bi--Conjugate Gradient (PBiCG) 
method with a DILU preconditioner was employed to solve the discretized 
equations for momentum $\nutilde$, $\gamma$ and $\Ret$ equations.
On the other hand, a Preconditioned Conjugate Gradient (PCG) solver with 
a diagonal incomplete Cholesky preconditioner was used for the pressure equation.\\
Finally, a local convergence tolerance of $10^{-6}$ was set for the pressure, 
while the other linear systems were considered converged once the residuals 
reached machine precision.
\subsection{Implementation aspects}
\noindent The \texttt{foam-extend} v5.0 package is an 
object--oriented C++ code and, as previously introduced, 
is a community-driven fork of the widely used OpenFOAM library.
These open--source libraries have attracted a significant
 portion of the CFD community, as their structure allows 
for the implementation of solvers and models with relatively 
low coding effort.\\
A notable feature of \texttt{foam-extend} and its derived libraries 
is the ability to closely mirror mathematical notation
in C++ code. In this context, two classes are particularly important:
\texttt{finiteVolumeCalculus} (or \texttt{fvc}), 
which performs an explicit evaluation of the tensorial operator; 
and \texttt{finiteVolumeMethod} (or \texttt{fvm}), which returns 
a matrix representation of the specific operation.\\
In our implementation strategy  wherever possible,
the involved the terms are treated implicitly using classes
related to \texttt{fvm}.
However, the cross terms, appearing in both $\gammatilde$ an SA equations, 
are implemented explicitly as
\texttt{foam-extend} does not provide built--in functions 
to  implicitly discretize similar non--linear terms.
In contrast, the destruction term SA equation has been linearized, since 
\texttt{fvm::Sp} (the high--level operator used to handle source terms)  
supports only linear terms.
Due to the lack of classes for exponential terms, the source terms in the 
intermittency equation are implemented fully explicitly.
It is also important to note that, in our strategy, 
both the production and destruction terms are clipped to zero when they change sign. 
This condition is not strictly necessary, since the terms are energy--consistent, 
as demonstrated above. Nevertheless, we enforce clipping to ensure the 
preservation of their intended physical behavior.\\
Lastly, we also want to emphasize that the $\Ret$ transport equation
 has been implemented in a standard form, as its terms do not 
require any special treatment or considerations.

\section{Results}\label{sec:res}
In this section, we discuss the predictive capabilities of 
the transition modelling approaches proposed in this paper.
Three different airfoils are selected as benchmark cases: SD7003 ($\mathrm{Re} = 6 \cdot 10^4$), 
E387 ($\mathrm{Re} = 2 \cdot 10^5$ and $\mathrm{Re} = 3 \cdot 10^5$), and DU00--W--212 
($\mathrm{Re} = 3 \cdot 10^6$ and $\mathrm{Re} = 6 \cdot 10^6$).
The aforementioned airfoils exhibit both the reliability of literature data available 
for the specific flow problem and a suitable level of complexity. 
Therefore, they can be considered appropriate choices to test our 
approach across a wide range $\mathrm{Re}$ numbers.\\
Moreover, \textcolor{black}{the following three flow models} will be included in the discussion
namely: $\log \gamma$--$\Ret$--SA--R1, $\log \gamma$--$\Ret$--SA--SMP, and 
a single version of $\log \gamma$--SA. 
It is important to remark that when a one--equation transition strategy is employed, only the SA--SMP 
formulation provides stable computations. Hence, for one--equation
approach, we intentionally omit specifying the SA equation version.\\
The numerical solutions presented below were obtained on a distributed--memory
 parallel Debian Linux cluster composed of four Intel Xeon 6238R nodes, 
totaling 192 CPU cores operating at 2.2 GHz. 
A typical simulation run was configured to use 24 CPU cores.
The \texttt{foam-extend} library was built using the GCC 10.2.1 compiler, 
and the MPI implementation used is MPICH 3.3.1, also built with GNU compilers.
\subsection{SD7003}
\begin{figure}[htbp]
 \centering
 {\includegraphics[width=0.45\textwidth]{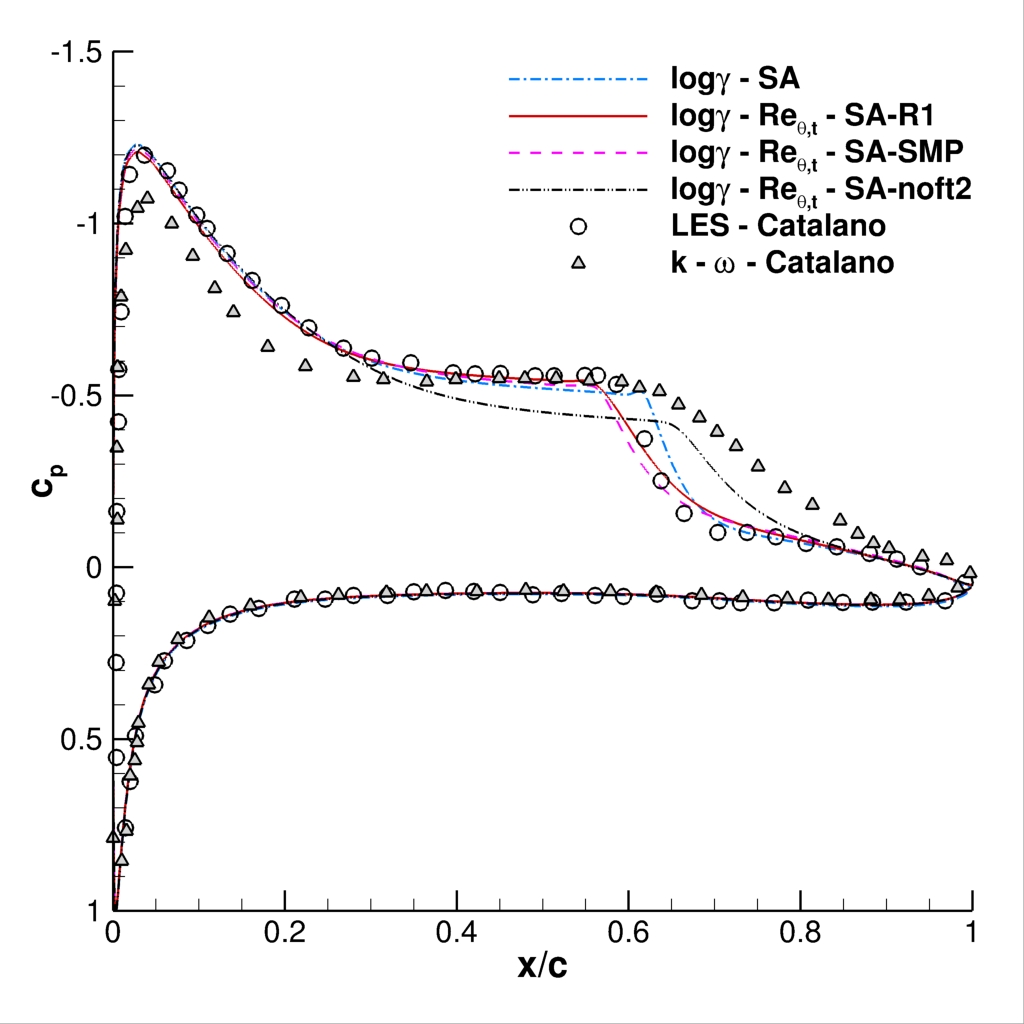}\label{fig:sd7003_cp_a4}}
        \caption{SD7003, $\mathrm{Re} = 6 \cdot 10^4$, $\alpha = 4^\circ$. Pressure coefficient distribution.}
\label{fig:sd7003_a4}
\end{figure}

\begin{figure}[htbp]
 \centering
 {\includegraphics[width=0.45\textwidth]{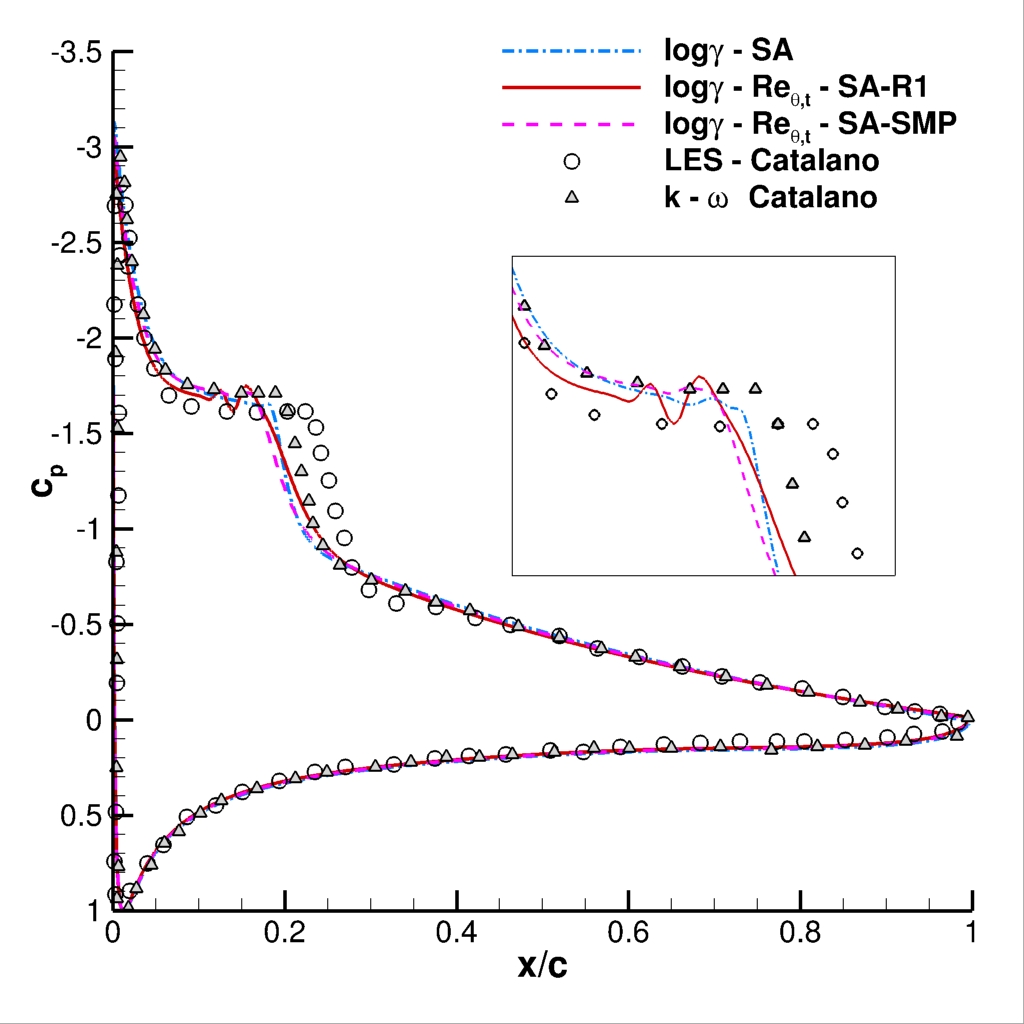}\label{fig:sd7003_cp_a8}}
        \caption{SD7003, $\mathrm{Re} = 6 \cdot 10^4$, $\alpha = 8^\circ$. Pressure coefficient distribution, baseline models.}
\label{fig:sd7003_a8}
\end{figure}

\begin{figure}[htbp]
 \centering
 {\includegraphics[width=0.45\textwidth]{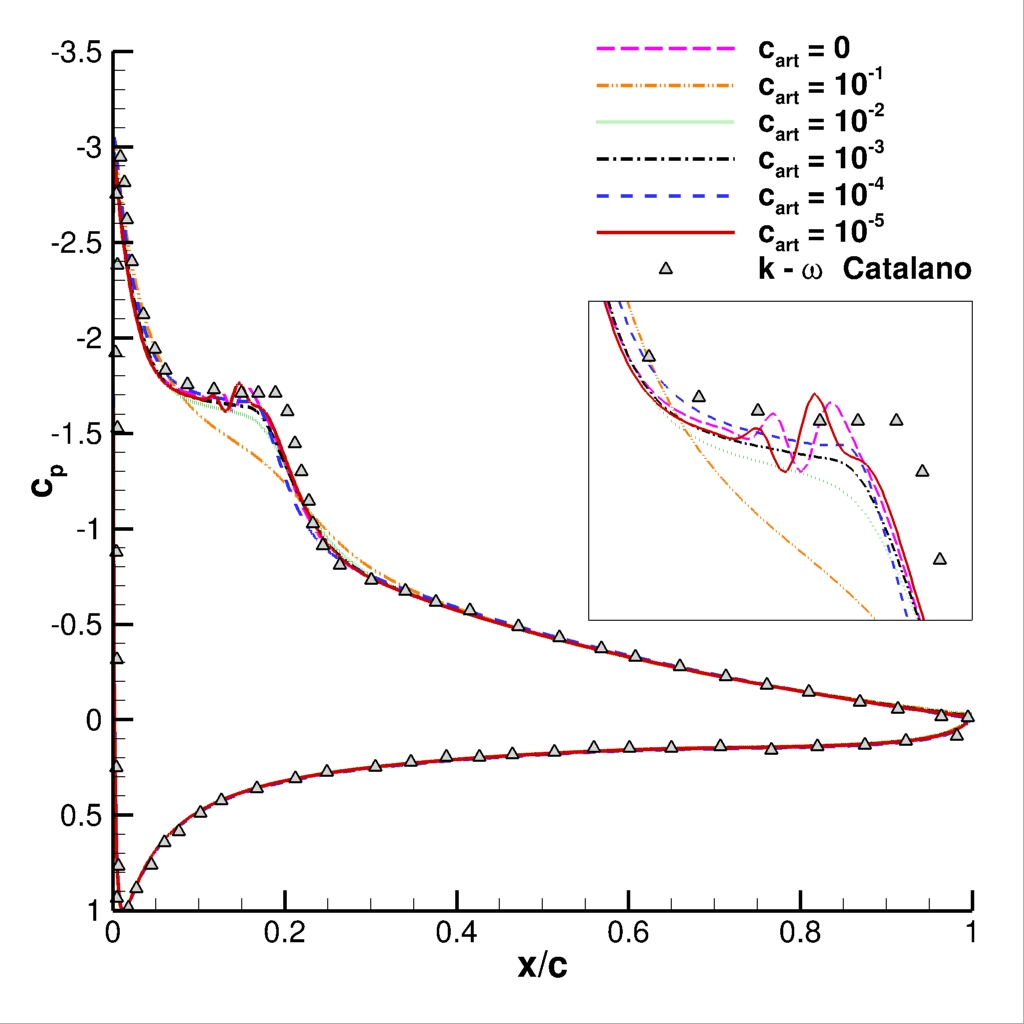}\label{fig:sd7003_cp_a8-r1-diff}}
        \caption{SD7003, $\mathrm{Re} = 6 \cdot 10^4$, $\alpha = 8^\circ$. Effect of artificial viscosity on pressure coefficient distribution, $\log \gamma$--$\Ret$--SA--R1.}
\label{fig:sd7003_a8-r1-diff}
\end{figure}

\begin{figure}[htbp]
 \centering
 {\includegraphics[width=0.45\textwidth]{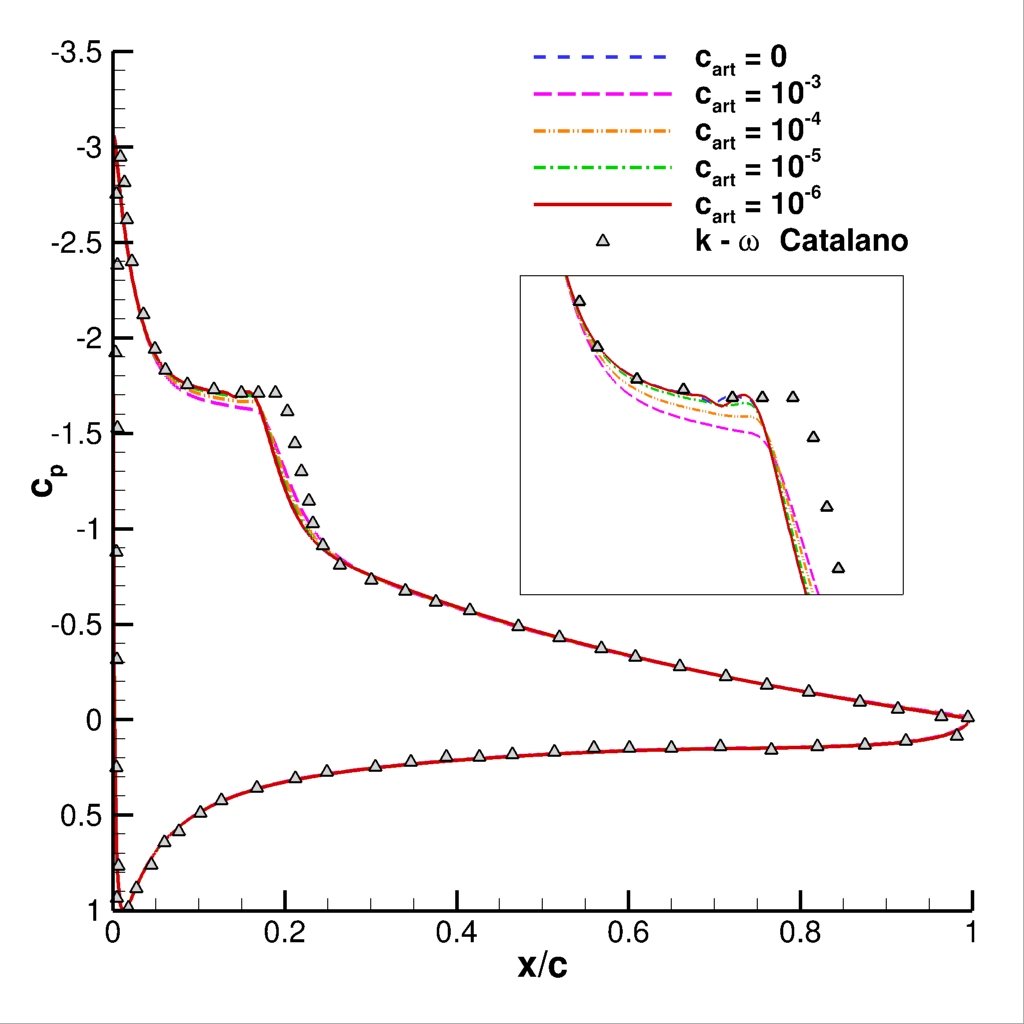}\label{fig:sd7003_cp_a8-SMP-diff}}
        \caption{SD7003, $\mathrm{Re} = 6 \cdot 10^4$, $\alpha = 8^\circ$. Effect of artificial viscosity on pressure coefficient distribution, $\log \gamma$--$\Ret$--SA--SMP.}
\label{fig:sd7003_a8-smp-diff}
\end{figure}

\begin{figure}[htbp]
 \centering
 {\includegraphics[width=0.45\textwidth]{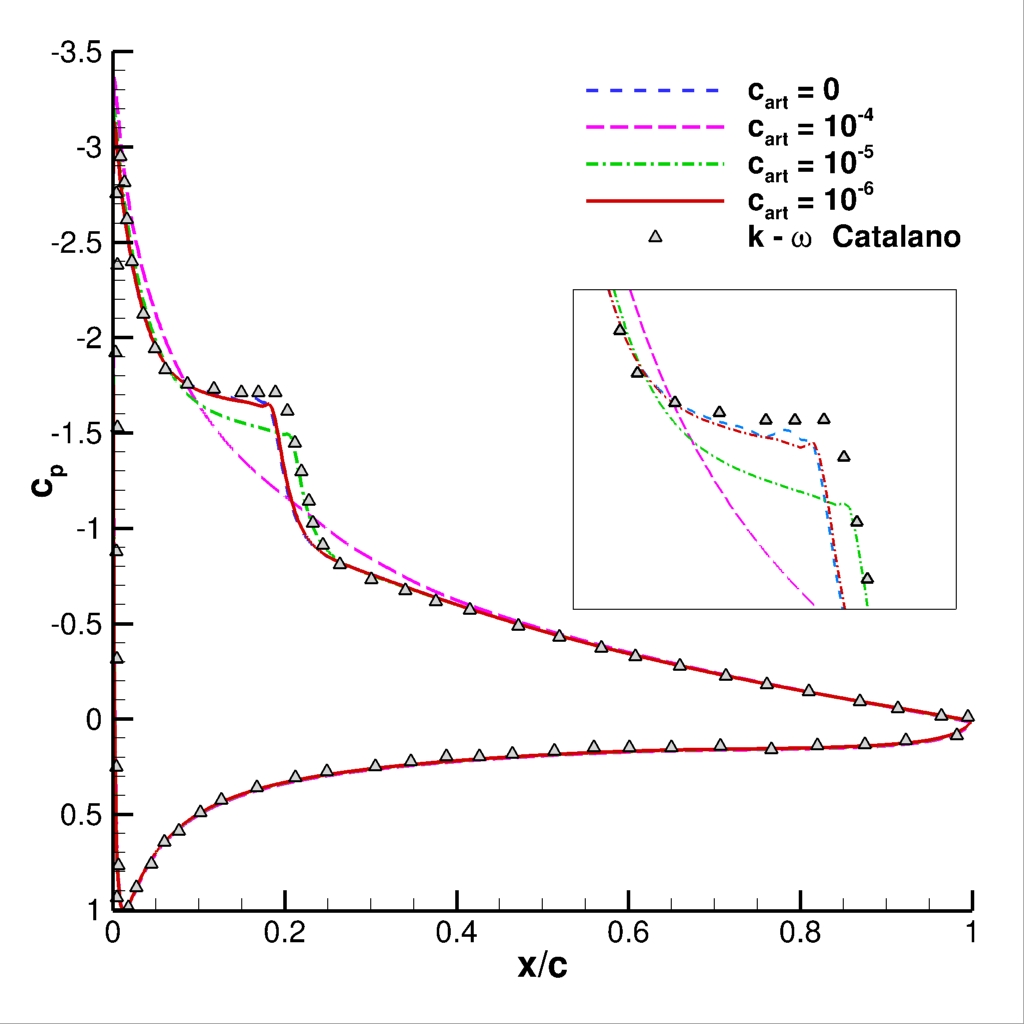}\label{fig:sd7003_cp_a8-1eq-diff}}
        \caption{SD7003, $\mathrm{Re} = 6 \cdot 10^4$, $\alpha = 8^\circ$. Effect of artificial viscosity on pressure coefficient distribution, $\log \gamma$--SA.}
\label{fig:sd7003_a8-1eq-diff}
\end{figure}

The first airfoil considered is the Selig–Donovan (SD) 7003, 
operating at $\mathrm{Re} = 6 \cdot 10^4$ with a free--stream
turbulence intensity of $0.05\%$.
This is a widely used test case in the literature~\cite{CATALANO2011615,Galbraith2008,Lian20071501,Uranga2011232}, 
as it exhibits significant variation in the size of the 
laminar separation bubble with changes in the angle of attack. 
Therefore, it represents a challenging scenario and has been selected 
to benchmark the model formulation presented in this paper.\\
Fig.~\ref{fig:sd7003_p_a4} shows the geometry of the airfoil 
and the surrounding dimensionless pressure field at $\alpha = 4^\circ$.
A C--topology grid with 768 cells in the streamwise direction 
(including 96 cells in the wake) and 176 cells in the
 wall--normal direction was used. 
This grid was generated at the CIRA (Italian Aerospace Research Center) and has 
been extensively tested for the computation of this type of 
flow field~\cite{CATALANO2011615}. 
\textcolor{black}{
On the other hand, Fig.~\ref{fig:sd7003_nut_nu} presents the contour 
plot of the turbulent--to--molecular viscosity ratio ($\nu_t/\nu$) 
along with velocity streamlines, clearly highlighting the presence of 
a laminar separation bubble. It is evident that flow separation 
occurs in the laminar region, where $\nu_t/\nu \rightarrow 0$. 
In contrast, the laminar--to--turbulent transition develops
 within the separated shear layer, where significant
 production of turbulent viscosity is observed. 
The reattachment point occurs in a region where the flow is fully turbulent.
}\\  
In the following, we compare the pressure coefficient distribution, defined as
 $c_p = 2\left( p - p_\infty\right)/\rho_\infty u_\infty^2$,
along the airfoil chord with the LES and $k$--$\omega$ results reported 
by Catalano and Tognaccini~\cite{Catalano_aiaa, CATALANO2011615}. 
At $\alpha=4^\circ$ the $c_p$ distribution shows
a good agreement between our results with LES data from and literature, 
see Fig.~\ref{fig:sd7003_a4}. 
By contrast, the $k$--$\omega$ results~\cite{CATALANO2011615} differ
 significantly from those obtained with LES.
Fig.~\ref{fig:sd7003_a4} also reports data obtained using the 
SA model version called \texttt{noft2},
\textcolor{black}{which has not been mentioned previously but
is widely used among various SA model implementations.}
In this particular case, the term $\Stilde$ in eq.~\ref{eq:clfun2} 
includes only the first terms inside the square brackets and does not 
include the second contribution related to the minimum condition.
The derived model formulation is a standard for the SA equation and is referred 
to as \texttt{noft2}, following the nomenclature adopted by the NASA 
Turbulence Model Benchmarking Working Group~\cite{NASA_TMR}.
It is easy to observe that the SA--noft2 formulation reliably predicts laminar separation, 
while the transition (the knee in the $c_p$--$x/c$ curve) is significantly delayed. 
This explains the large discrepancy with LES data for $x/c > 60\%$. 
For this reason, we consider this formulation unsuitable for the flow regime of 
interest in this paper and therefore exclude it from the following analysis.
This result is particularly relevant in the context of the present work, as 
it highlights that the turbulent transition in the separated shear layer 
is better captured by the $\tilde{S}$ terms employed in the SA--R1 
and SA--SMP formulations.\\
Results for $\alpha=8^\circ$ are shown in Fig.~\ref{fig:sd7003_a8}.
The predicted $c_p$ distributions from all models follow the same general trend.
In this case, the transition is slightly underpredicted compared to LES data,
and our results show better agreement with the $k$--$\omega$ model, particularly
in the post transition region.
Nevertheless, a critical issue is the Gibbs--like phenomenon exhibited by all the models
near the pressure knee. This effect is particularly pronounced in the SA--R1 version.
These spurious pressure fluctuations are  evident in the transitional region 
and significantly compromise the accuracy of the pressure coefficient.
In order to mitigate this behaviour the gradient--driven artificial viscosity,
discussed in Sec.~\ref{sec:goveq}, was activated. 
Fig.~\ref{fig:sd7003_a8-r1-diff} shows the sensitivity of the SA--R1 variant to the artificial viscosity
scaling coefficient $c_{art}$.
As expected, large $c_{art}$ values introduce excessive dissipation, which can
suppress the laminar separation bubble.
On the other hand, if the dissipation is carefully calibrated then the viscosity 
selectively damps pressure wiggles.
In this case, a value of $c_{art} = 10^{-4}$ was found to provide the best compromise,
since it dissipates suitably pressure oscillations maintaining consistency with the
 physically expected solution.\\
Fig.~\ref{fig:sd7003_a8-smp-diff} and Fig.~\ref{fig:sd7003_a8-1eq-diff} show the 
sensitivity of the SA--SMP model and the one--equation with respect to $c_{art}$
coefficient.
The overall behaviour of these models is qualitatively similar to SA--R1 approach.
However, a key difference lies in the magnitude of $c_{art}$ required to achieve 
a physically consistent solution.
More precisely, the SA--SMP model achieves optimal smoothing with $c_{art} = 10^{-5}$,
 whereas the one--equation model requires $c_{art} = 10^{-6}$. 
%
These observations underscore that the three flow models examined require different levels of 
artificial viscosity to achieve numerical stability and smoothness. 
In particular, the SMP variant of the SA equation, as well as the one--equation formulation for transition, 
tend to introduce a higher degree of numerical diffusion.\\
%
%
%
\begin{figure}[htbp]
 \centering
 {\includegraphics[width=0.45\textwidth]{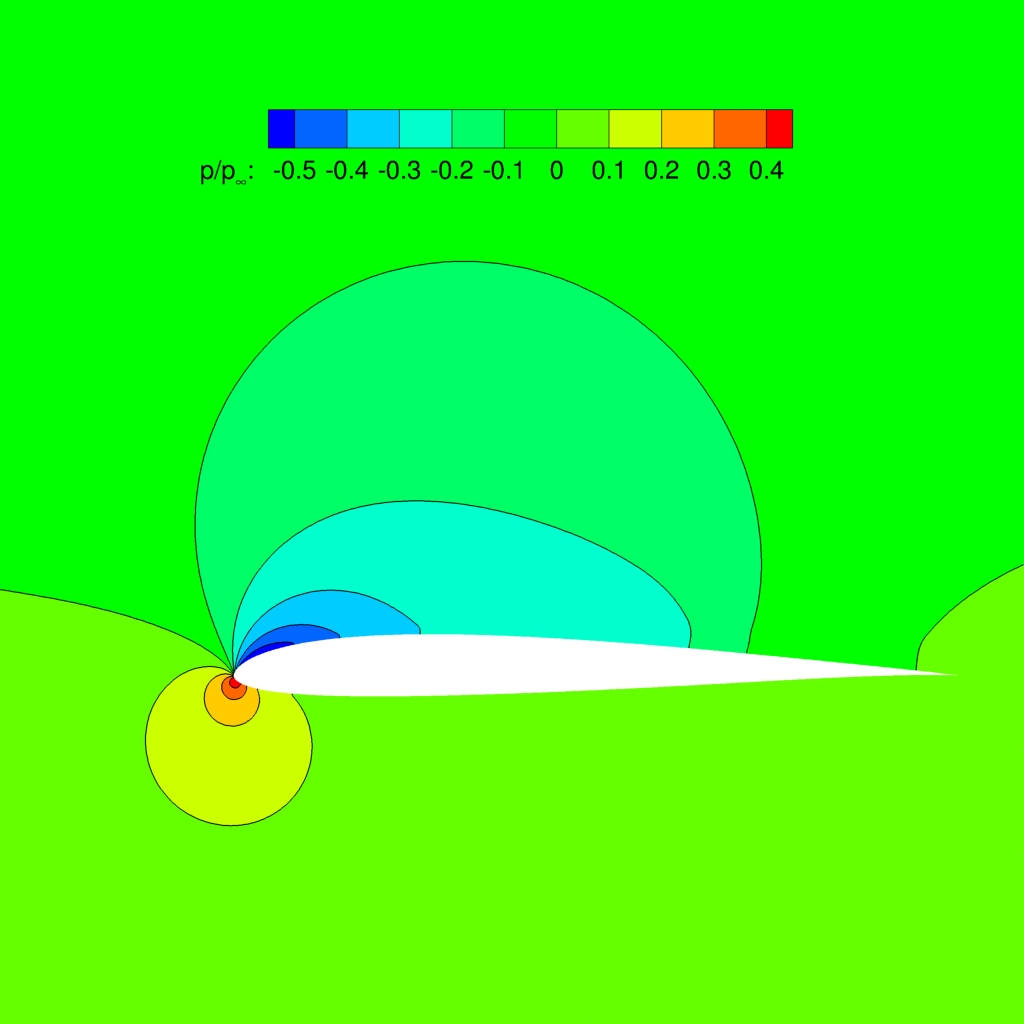}\label{fig:sd7003_p_a4}}
        \caption{SD7003, $\mathrm{Re} = 6 \cdot 10^4$, $\alpha = 4^\circ$. Dimensionless pressure field, SA--R1 model.}
\label{fig:sd7003_p_a4}
\end{figure}
\begin{figure}[htbp]
 \centering
 {\includegraphics[width=0.45\textwidth]{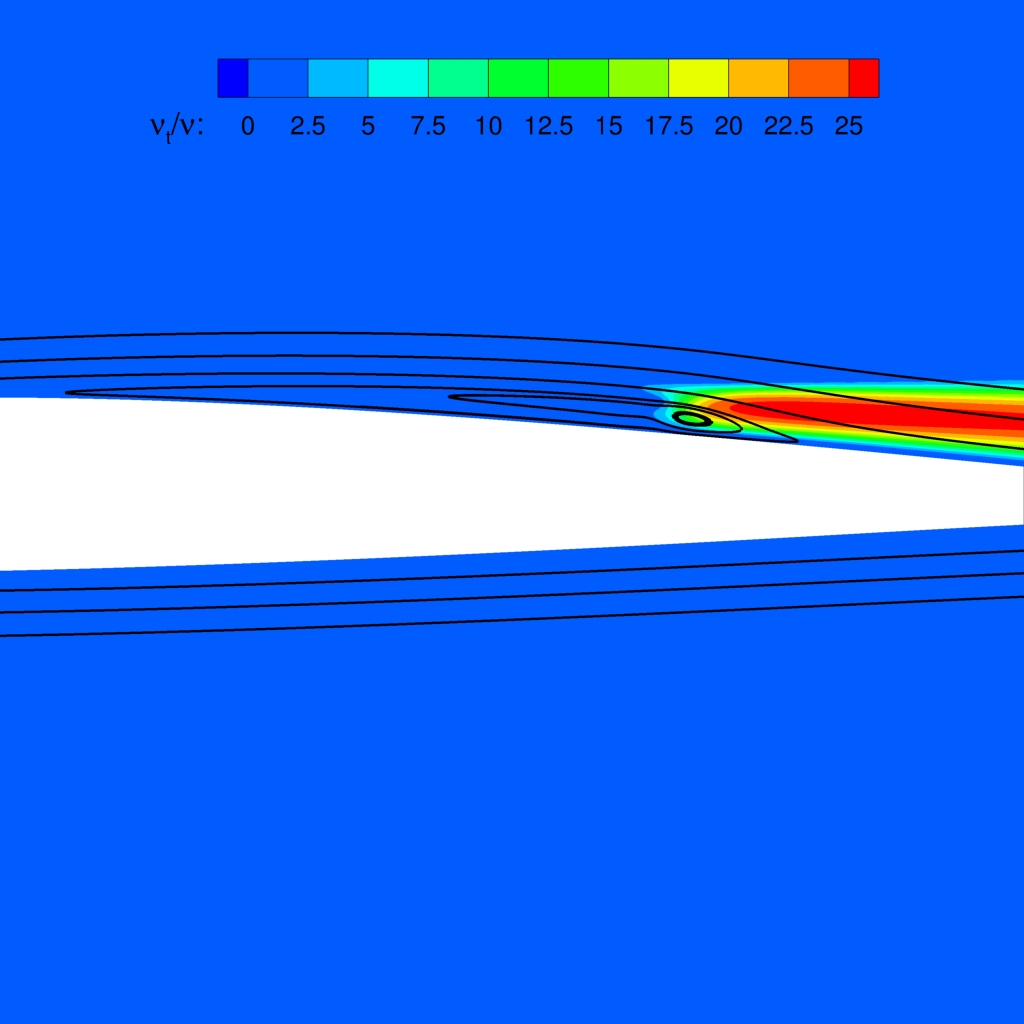}\label{fig:sd7003_nut_nu}}
        \caption{SD7003, $\mathrm{Re} = 6 \cdot 10^4$, $\alpha = 4^\circ$. SA--R1 model. $\nu_t/\nu$ contour plot and 
                 streamlines.}
\label{fig:sd7003_nut_nu}
\end{figure}
%
%
%
%
%
%
%
%
%
%
\begin{figure}[htbp]
 \centering
 {\includegraphics[width=0.45\textwidth]{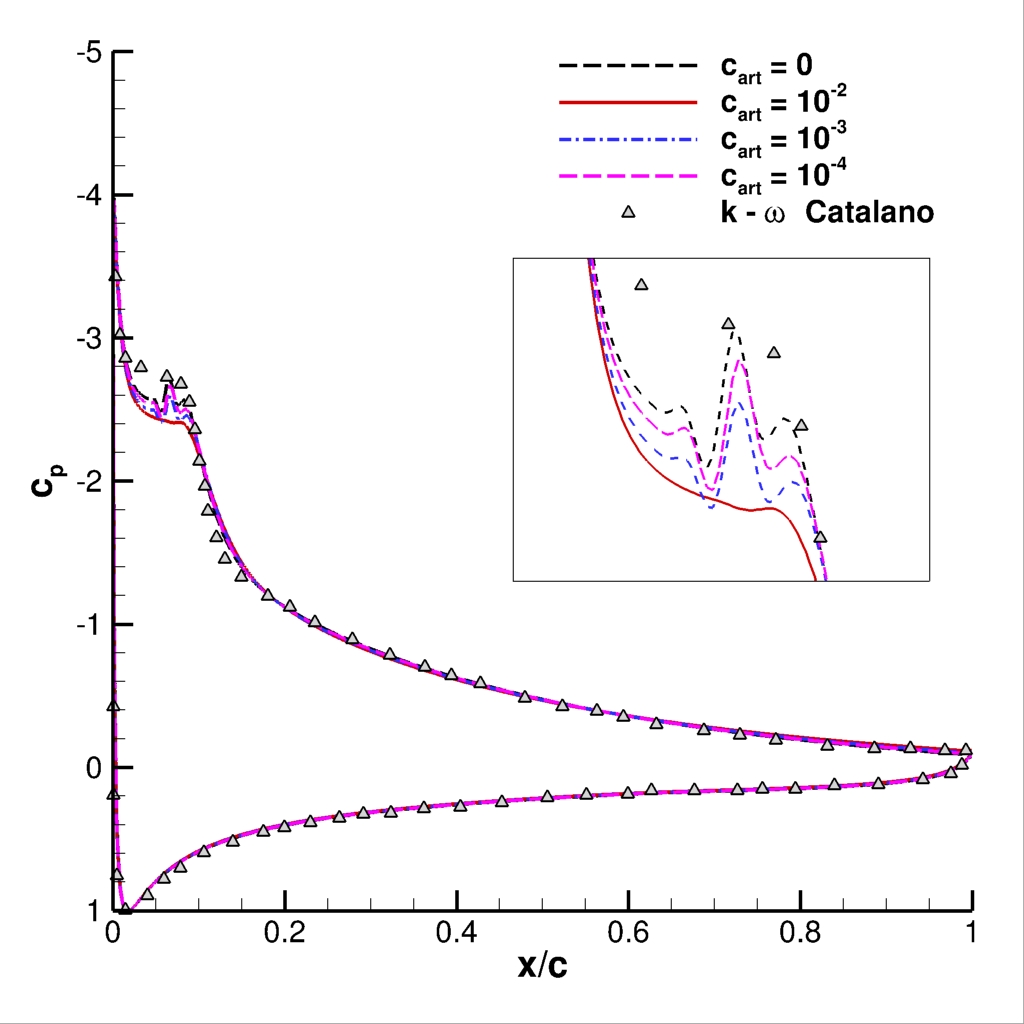}\label{fig:sd7003_cp_a10-SMP-diff}}
        \caption{SD7003, $\mathrm{Re} = 6 \cdot 10^4$, $\alpha = 10^\circ$. Effect of artificial viscosity on pressure coefficient distribution, $\log \gamma$--$\Ret$--SA--SMP.}
\label{fig:sd7003_a10-smp-diff}
\end{figure}
%
%
\begin{figure}[htbp]
 \centering
 {\includegraphics[width=0.45\textwidth]{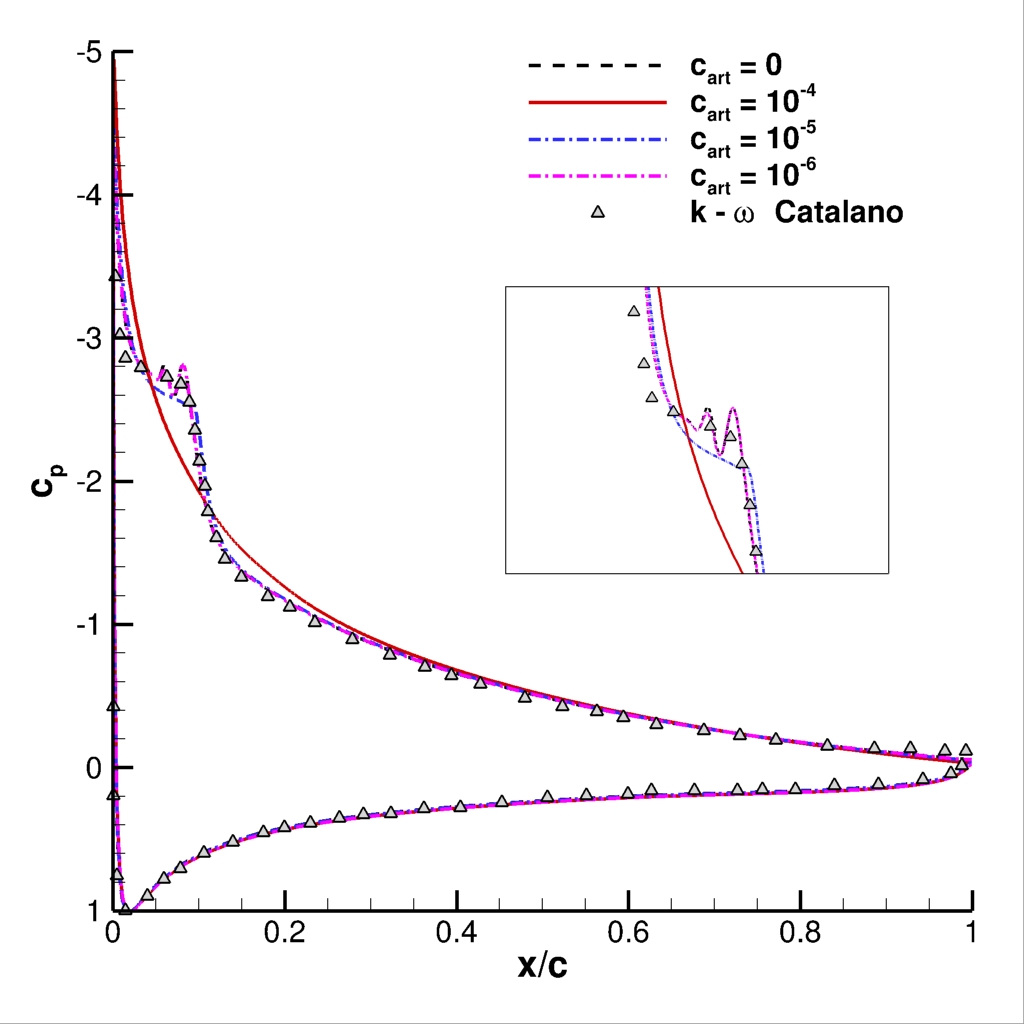}\label{fig:sd7003_cp_a10-1eq-diff}}
        \caption{SD7003, $\mathrm{Re} = 6 \cdot 10^4$, $\alpha = 10^\circ$. Effect of artificial viscosity on pressure coefficient distribution, $\log \gamma$--SA.}
\label{fig:sd7003_a10-1eq-diff}
\end{figure}
%
%
\begin{figure}[htbp]
 \centering
 {\includegraphics[width=0.45\textwidth]{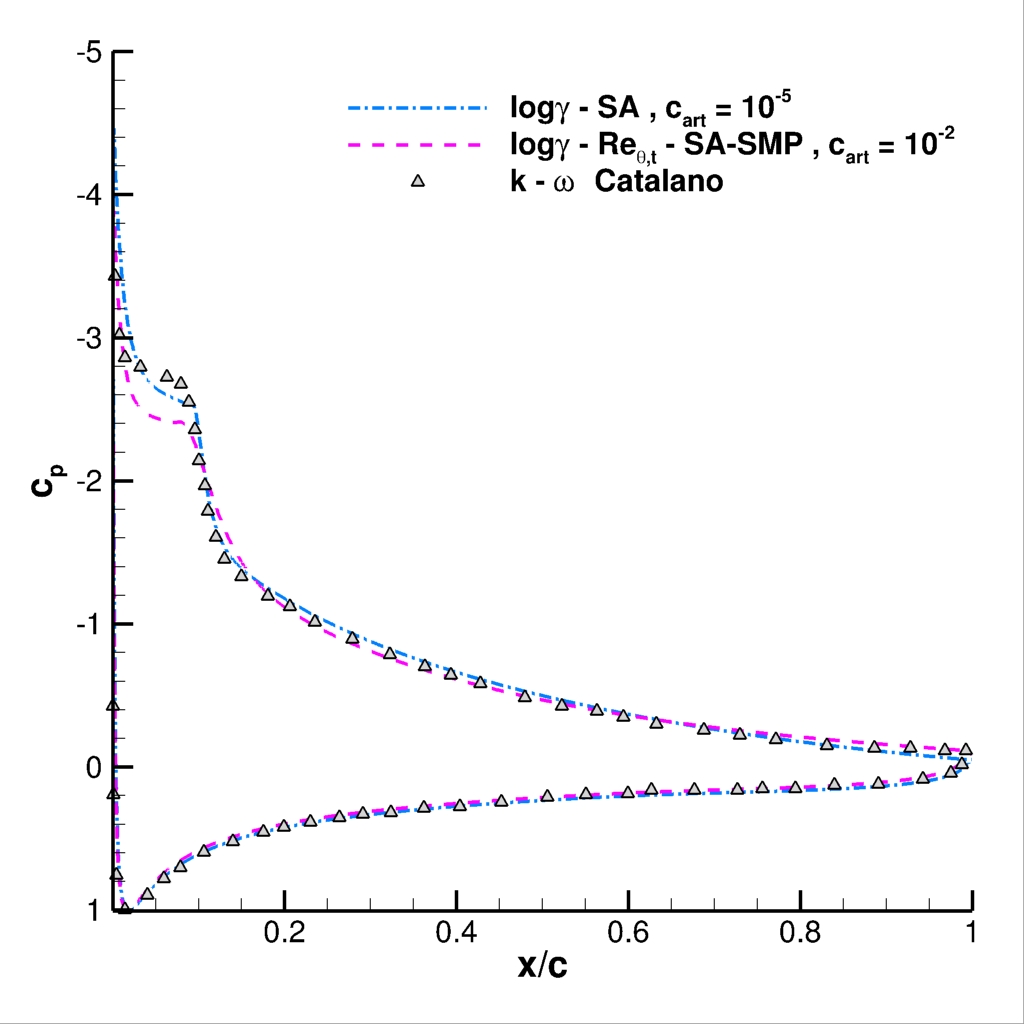}\label{fig:sd7003_cp_a10-comp}}
        \caption{SD7003, $\mathrm{Re} = 6 \cdot 10^4$, $\alpha = 10^\circ$. Effect of artificial viscosity on pressure coefficient distribution, models comparison.}
\label{fig:sd7003_a10-comp}
\end{figure}
%
\begin{figure}[htbp]
 \centering
 \includegraphics[width=0.45\textwidth]{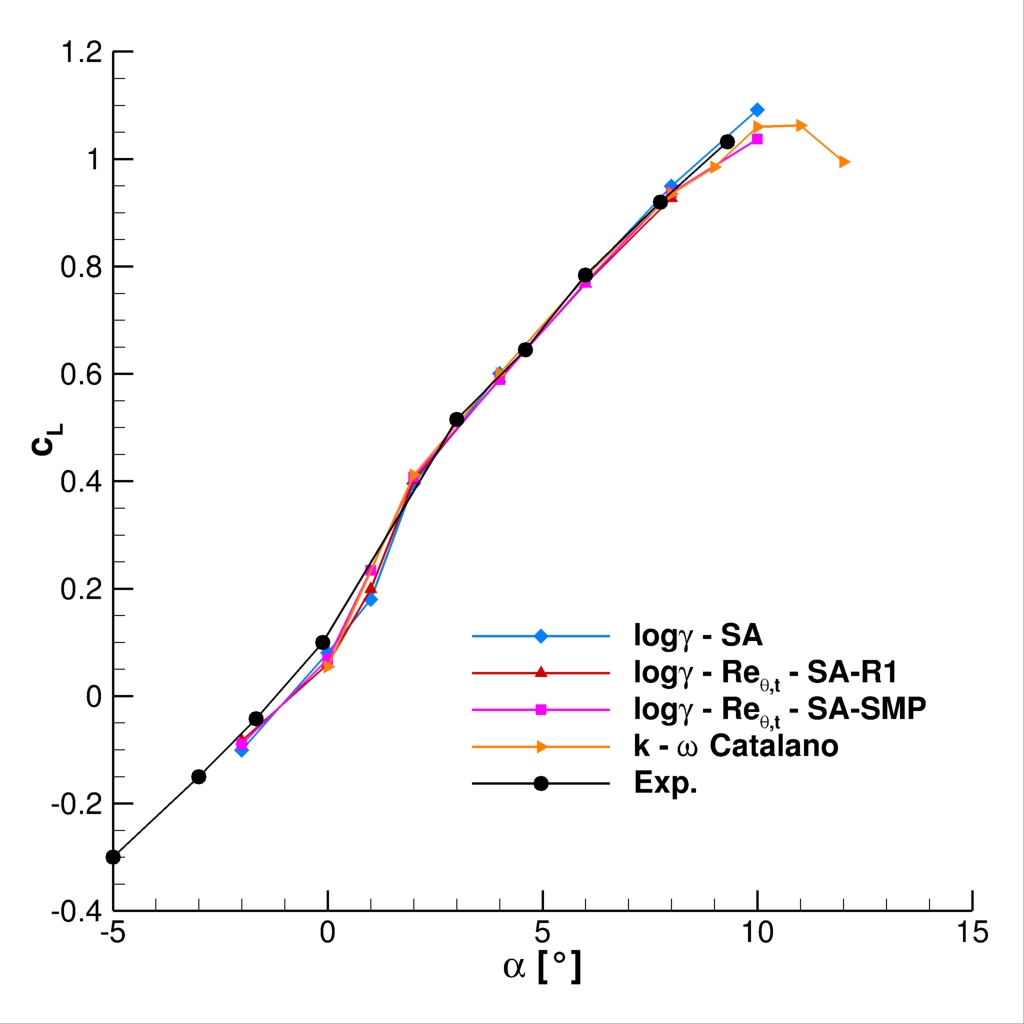}\label{fig:sd7003_cl-alpha}
        \caption{SD7003, $\mathrm{Re} = 6 \cdot 10^4$. Lift coefficient.}
\label{fig:sd7003_cl-alpha}
\end{figure}

\hspace{-0.75cm} At $\alpha=10^\circ$ the pressure oscillations are even more evident than at $\alpha=8^\circ$
for both $\log \gamma$--$\Ret$--SA--SMP, and $\log \gamma$--SA models.
In contrast, for the $\log \gamma$--$\Ret$--SA--R1 configuration steady--state solution was not obtained.
Fig.~\ref{fig:sd7003_a10-smp-diff} and Fig.~\ref{fig:sd7003_a10-1eq-diff}  
show the sensitivity with respect to $c_{art}$ coefficient. It is worth noting 
that the trends observed for $\alpha=8^\circ$ are confirmed. 
In particular, one--equation model variant requires a lower amount of artificial viscosity
to damp spurious pressure oscillations.
Finally, Fig.~\ref{fig:sd7003_a10-comp} presents a comparison of the smoothed solutions for $\alpha=10^\circ$.
Of particular interest is the fact that the one--equation model shows
a  good agreement with reference $k$--$\omega$ data from the literature, whereas two--equation technique
clearly underestimates the pressure plateau.\\
The force coefficients are shown, in Fig.~\ref{fig:sd7003_cl-alpha} and Fig.~\ref{fig:sd7003_cl-cd} 
 with both numerical results~\cite{CATALANO2011615} and experimental data~\cite{Selig:1995}.
For comparison, the lift coefficient, $C_L = 2L'/\rho_\infty u_\infty^2 c$, 
and drag coefficient,  $C_D = 2D'/\rho_\infty u_\infty^2 c$ are considered.
\begin{figure}[htbp]
 \centering
 {\includegraphics[width=0.45\textwidth]{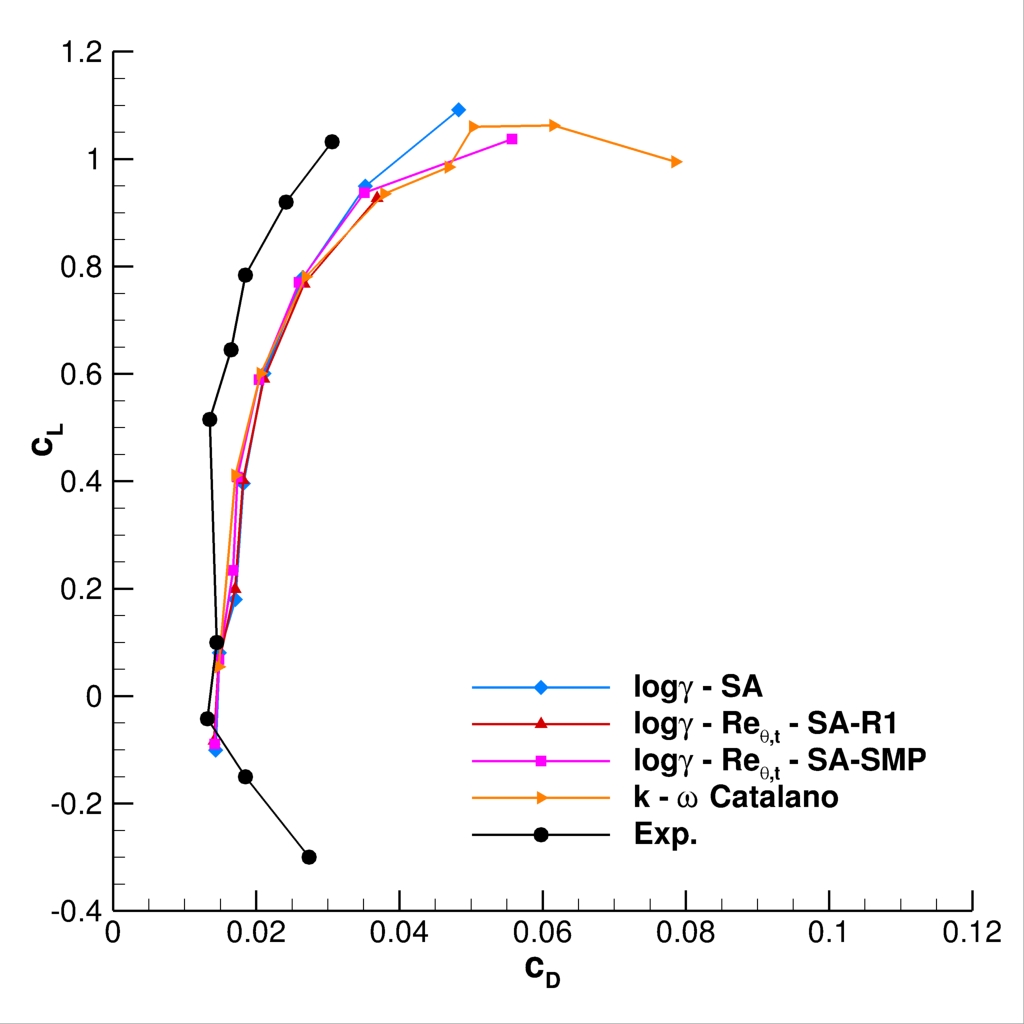}\label{fig:sd7003_cl-cd}}
        \caption{SD7003, $\mathrm{Re} = 6 \cdot 10^4$. Eiffel polar.}
\label{fig:sd7003_cl-cd}
\end{figure}
At low to moderate angles of attack, the computed lift coefficients 
show good agreement with the reference data. However, stall behavior 
is predicted differently by the models presented in this work. 
In particular, for this specific case, steady solutions are obtained 
only up to $\alpha = 10^{\circ}$. As a result, the model's performance in the stall
 region is fairly limited.
Conversely, for the drag coefficient, our model shows excellent agreement 
with numerical reference data prior to stall, while the results highlight an 
overestimation when compared to experimental data.\\
Note that force coefficients data, shown 
in Figs.~\ref{fig:sd7003_cl-alpha} and~\ref{fig:sd7003_cl-cd}, for the highest 
angles of attack were obtained using suitably selected $c_{art}$ value to suppress
nonphysical pressure oscillations.
\subsection{Eppler 387}
\begin{figure}[htbp]
 \centering
 {\includegraphics[width=0.45\textwidth]{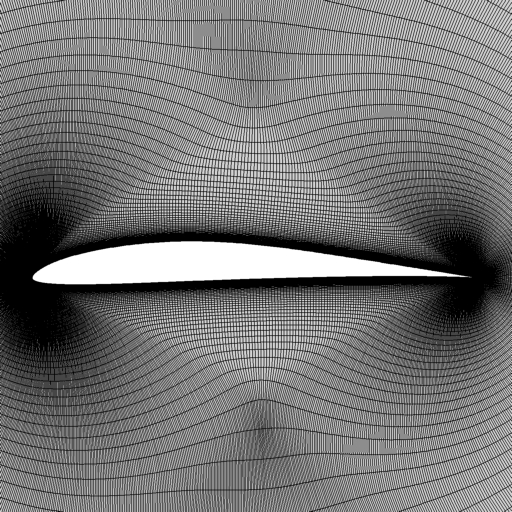}\label{fig:e387}}
        \caption{Computational grid around E387 airfoil.}
\label{fig:e387-grid}
\end{figure}

\begin{figure}[htbp]
 \centering
 {\includegraphics[width=0.45\textwidth]{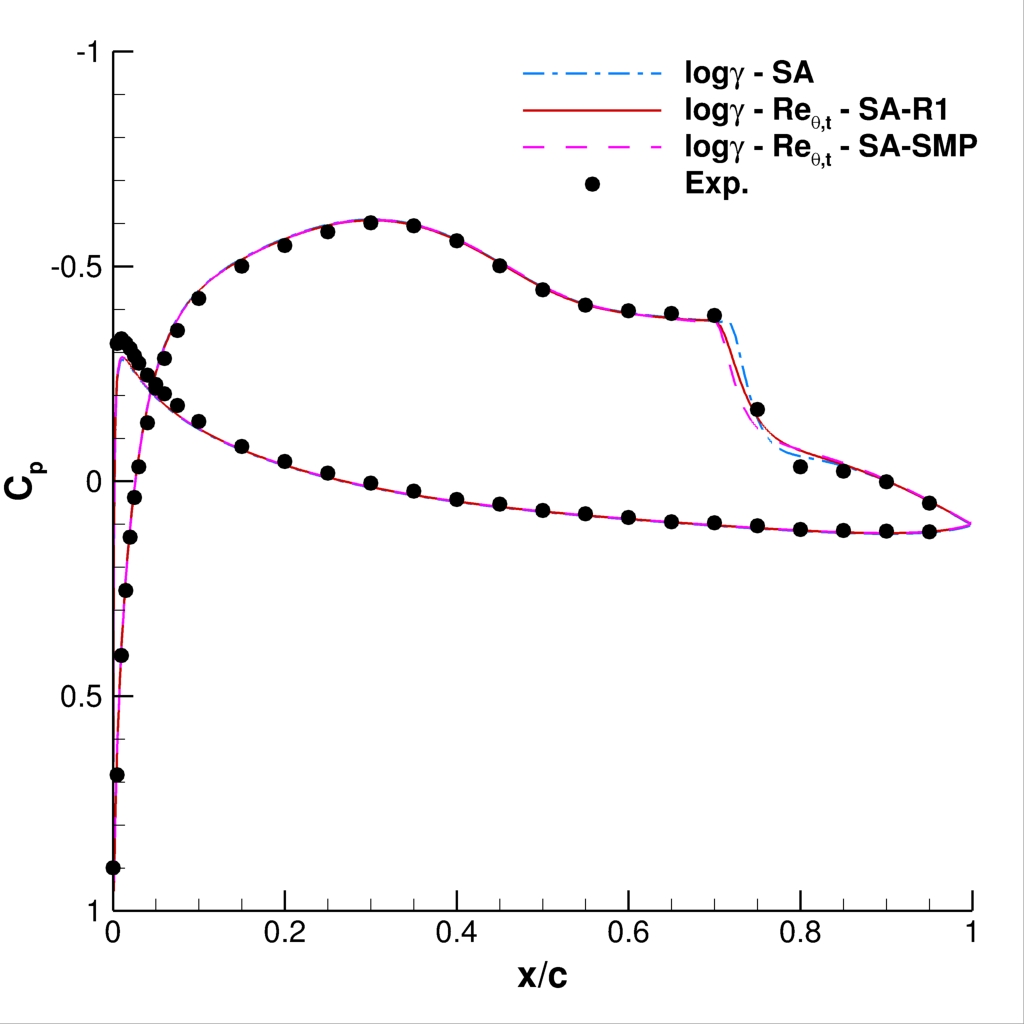}\label{fig:e387_cp_a0}}
        \caption{E387, $\mathrm{Re} = 2\cdot 10^5$, $\alpha = 0^\circ$. Pressure coefficient distribution.}
\label{fig:e387_a0}
\end{figure}

\begin{figure}[htbp]
 \centering
 {\includegraphics[width=0.45\textwidth]{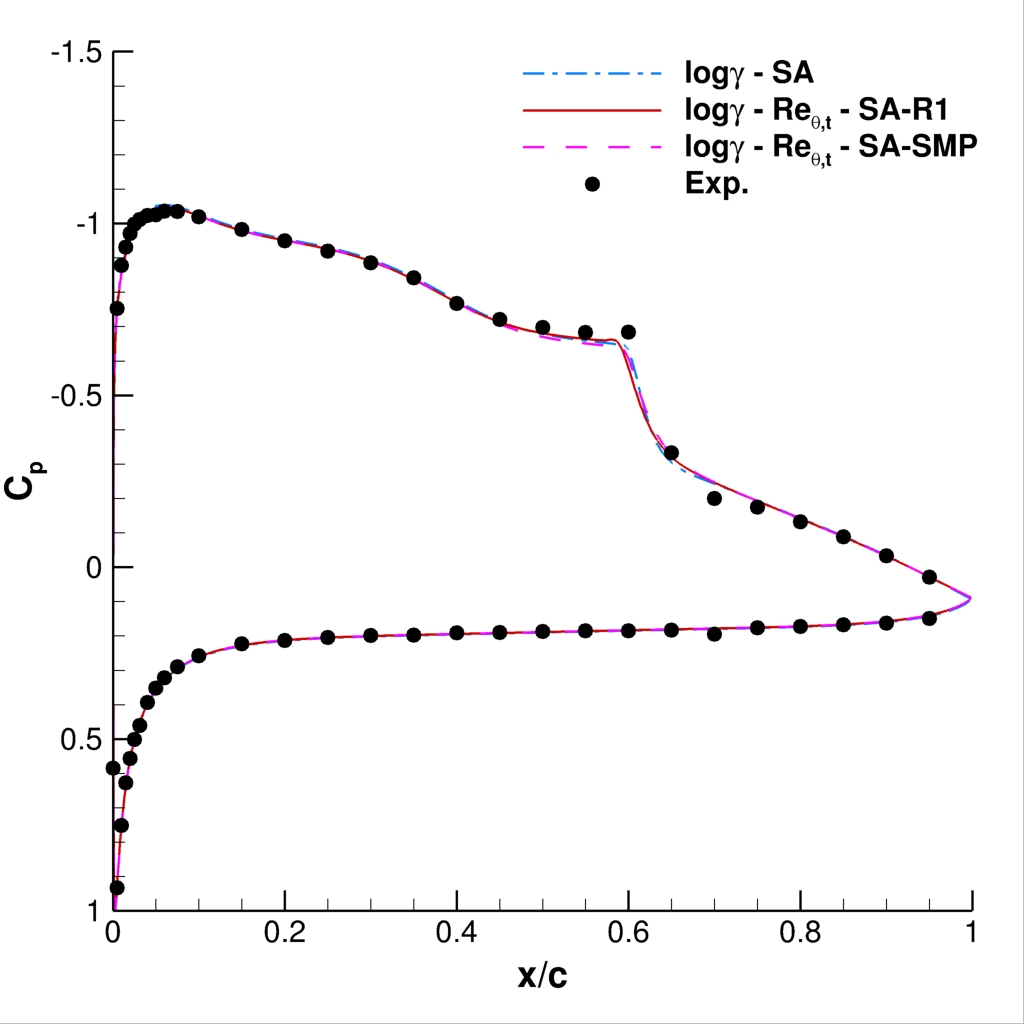}\label{fig:e387_cp_a4}}
        \caption{E387, $\mathrm{Re} = 2\cdot 10^5$, $\alpha = 4^\circ$. Pressure coefficient distribution.}
\label{fig:e387_a4}
\end{figure}

\begin{figure}[htbp]
 \centering
 {\includegraphics[width=0.45\textwidth]{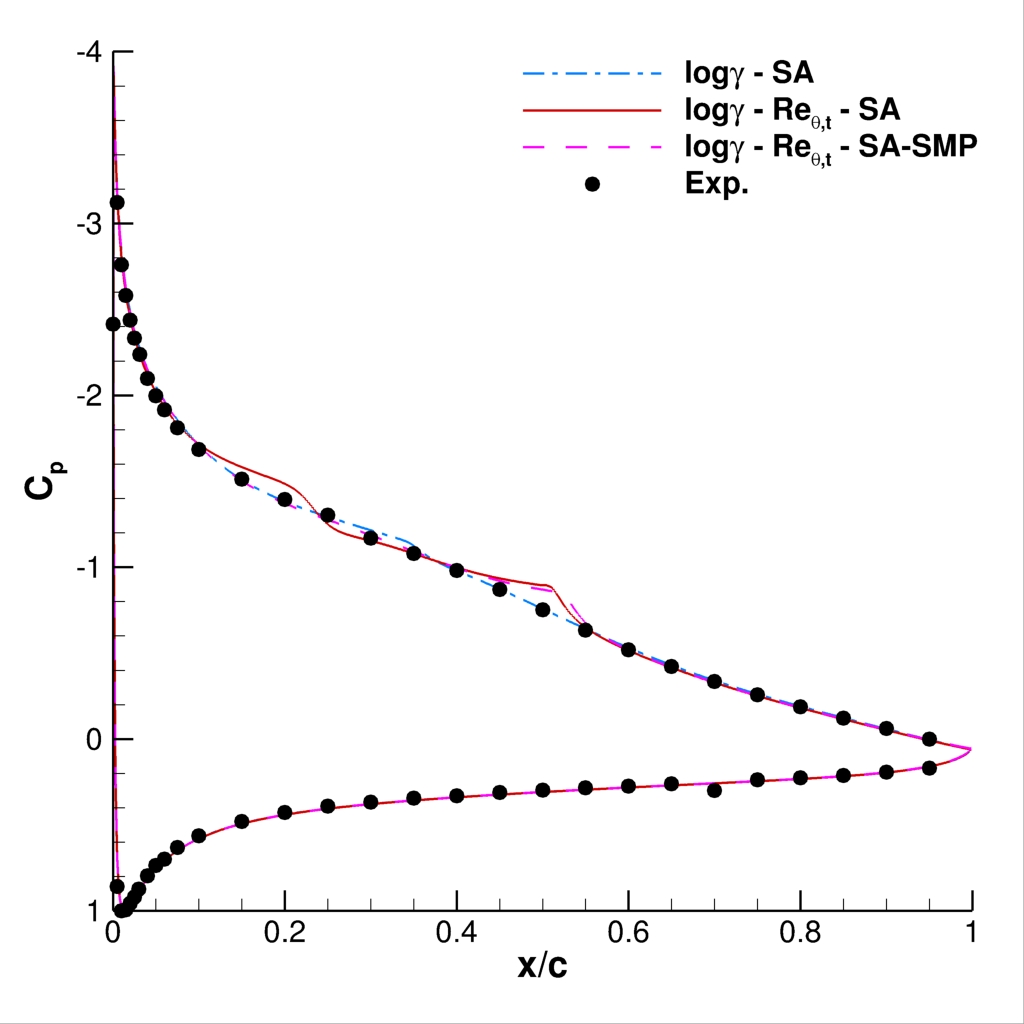}\label{fig:e387_cp_a8}}
        \caption{E387, $\mathrm{Re} = 2\cdot 10^5$, $\alpha = 8^\circ$. Pressure coefficient distribution.}
\label{fig:e387_a8}
\end{figure}

\begin{figure}[htbp]
 \centering
 {\includegraphics[width=0.45\textwidth]{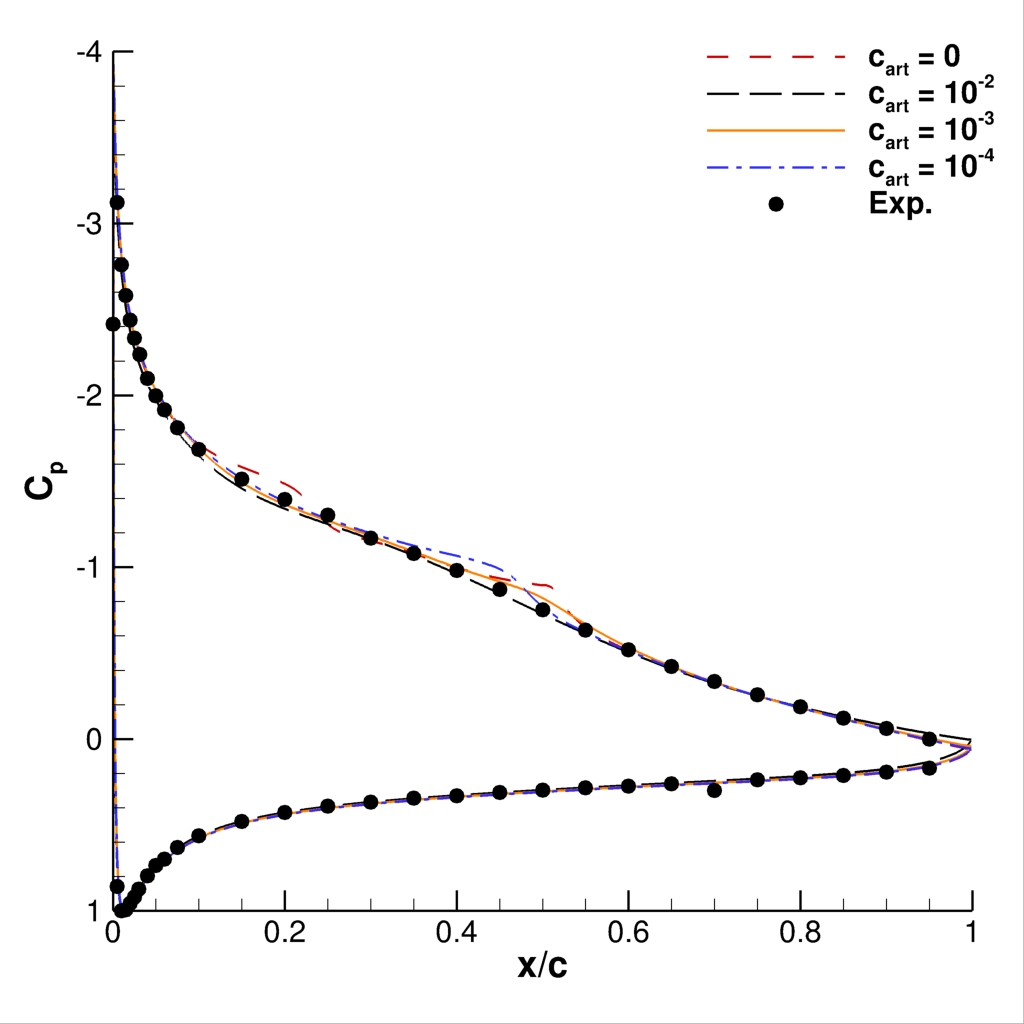}\label{fig:e387_cp_a8-diff}}
        \caption{E387, $\mathrm{Re} = 2\cdot 10^5$, $\alpha = 8^\circ$. Effect of artificial viscosity on pressure coefficient distribution, $\log \gamma$--$\Ret$--SA--R1.}
\label{fig:e387_a8-diff}
\end{figure}

\begin{figure}[htbp]
 \centering
 {\includegraphics[width=0.45\textwidth]{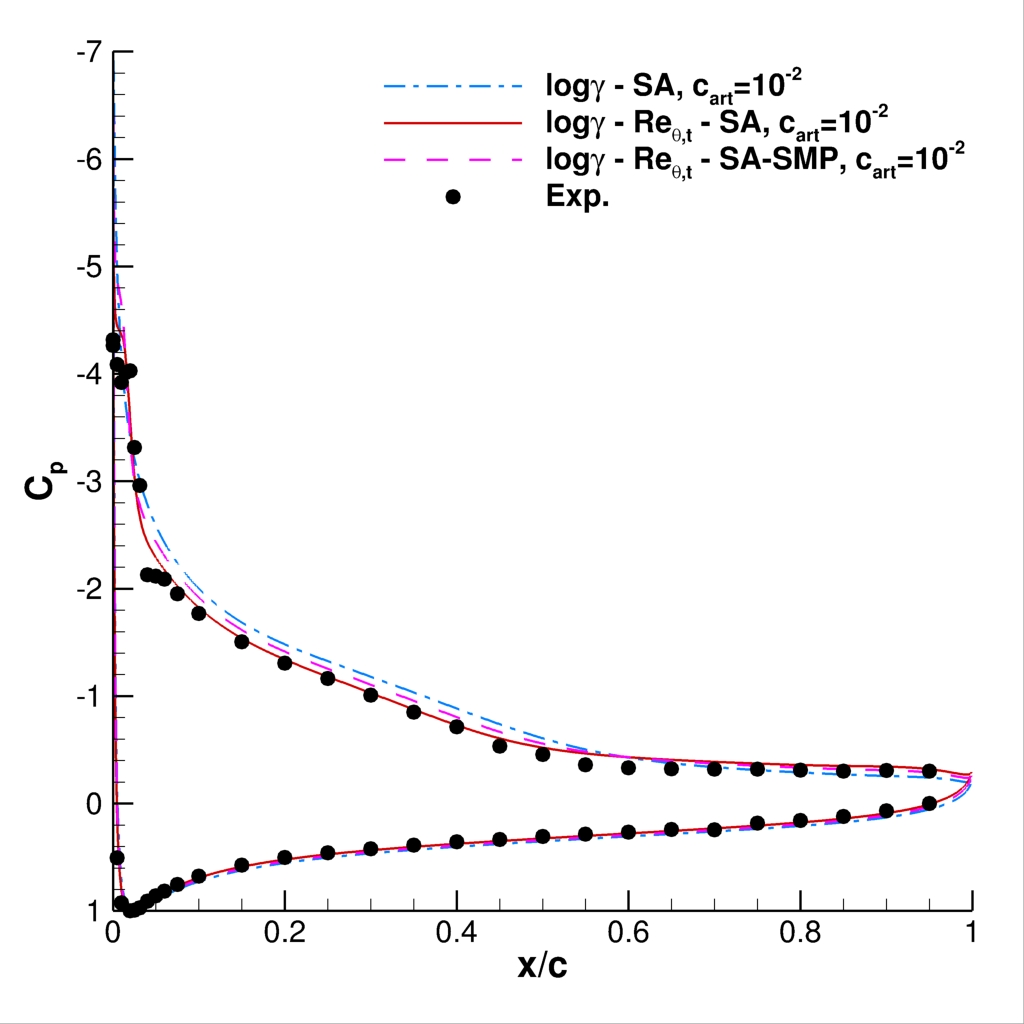}\label{fig:e387_cp_a12}}
        \caption{E387, $\mathrm{Re} = 2\cdot 10^5$, $\alpha = 12^\circ$. Pressure coefficient distribution.}
\label{fig:e387_a12}
\end{figure}

\begin{figure}[htbp]
 \centering
 {\includegraphics[width=0.45\textwidth]{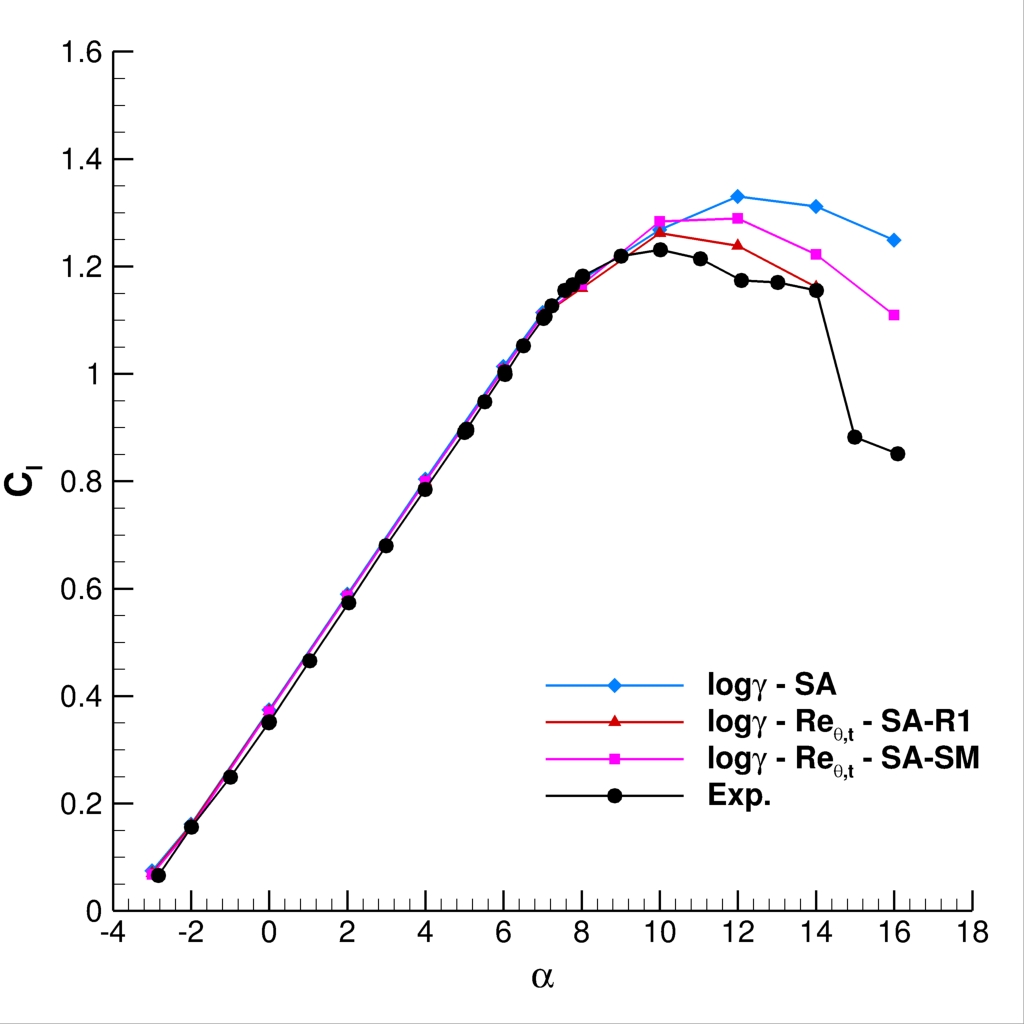}\label{fig:e387_cl}}
        \caption{E387, $\mathrm{Re} = 2 \cdot 10^5$. Lift coefficient.}
\label{fig:e387_cl-alpha-re200k}
\end{figure}

\begin{figure}[htbp]
 \centering
 {\includegraphics[width=0.45\textwidth]{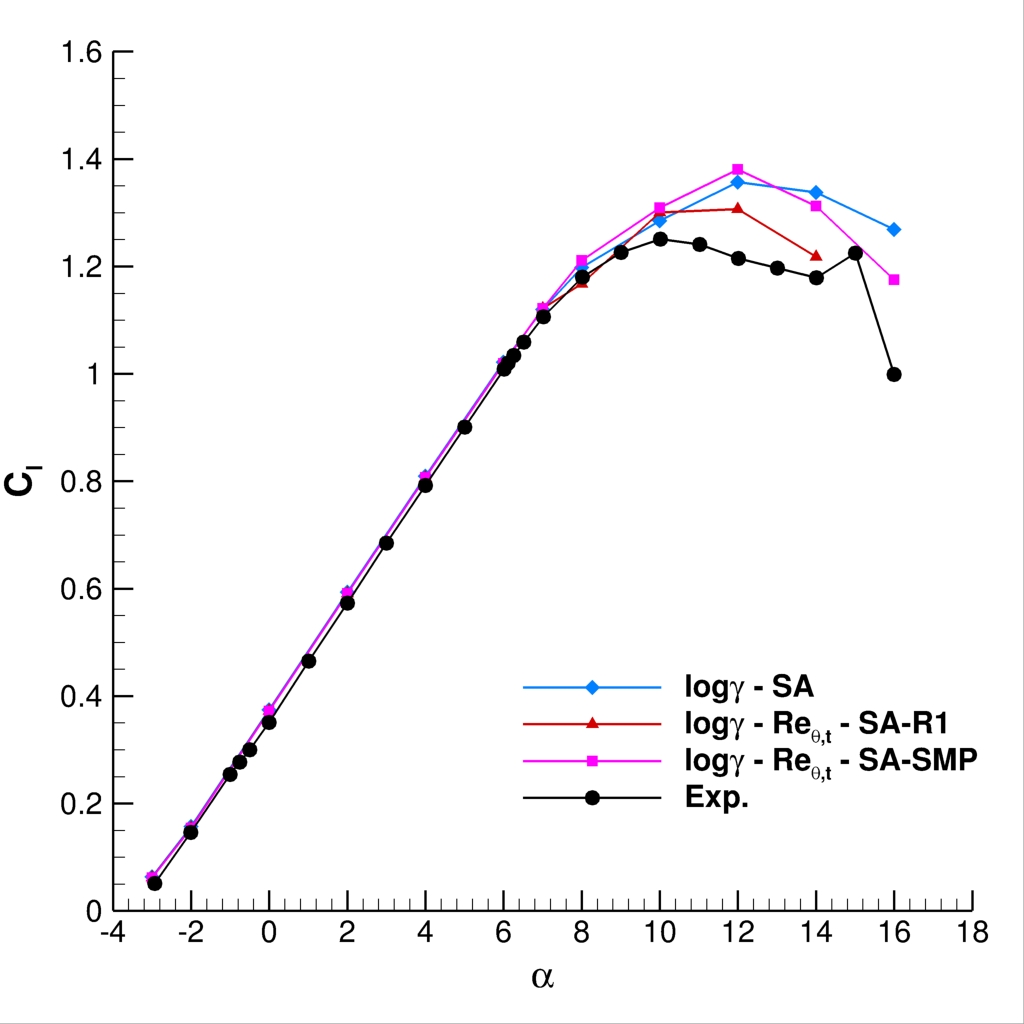}\label{fig:e387_cl}}
        \caption{E387, $\mathrm{Re} = 3 \cdot 10^5$. Lift coefficient.}
\label{fig:e387_cl-alpha-re300k}
\end{figure}

\begin{figure}[htbp]
 \centering
 {\includegraphics[width=0.45\textwidth]{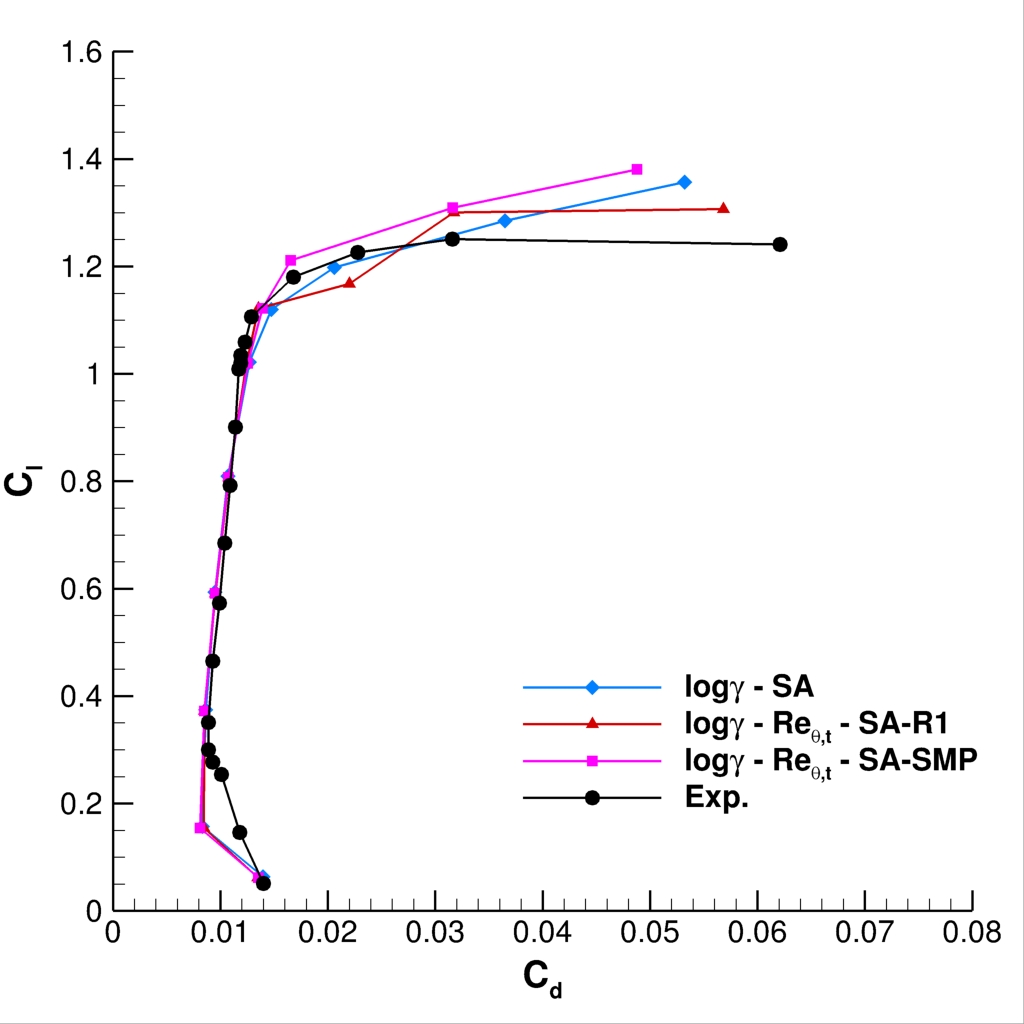}\label{fig:e387_cl-cd}}
        \caption{E387, $\mathrm{Re} = 3 \cdot 10^5$. Eiffel polar.}
\label{fig:e387_cl-cd-re300k}
\end{figure}

In this subsection, we present the obtained results for the flow 
over an Eppler 387 (E387) airfoil; see Fig.\ref{fig:e387-grid} for a representation.
The E387 airfoil is particularly well suited for this study due to 
the availability of a comprehensive experimental dataset published by 
McGhee et al.~\cite{McGhee:1988}, 
which includes detailed surface pressure distributions and force coefficients 
across a range of angles of attack and Reynolds numbers.\\
The numerical simulations were carried out using a structured O--type grid consisting 
of approximately $3 \cdot 10^5$ cells. The chord--based 
$\mathrm{Re}$ numbers considered are $2\cdot 10^5$ and $3\cdot 10^5$, with a free--stream turbulence of
$0.07\%$ in both the cases.
The inflow and outflow boundaries were placed at a distance of approximately 120 chord 
lengths away from the airfoil.
The grid was refined in the near--wall region, 
with the height of the first cell set to achieve $O\left( y^+\right) \simeq 1$.
Figs.~\ref{fig:e387_a0} to~\ref{fig:e387_a12} show the 
pressure coefficient distributions for
$\alpha = 0^\circ, 4^\circ, 8^\circ$, and $12^\circ$ 
at $\mathrm{Re} = 2 \cdot 10^5$. Other angles of attack, 
as well as the case at $\mathrm{Re} = 3 \cdot 10^5$, 
are omitted for the sake of compactness.
The reported results show good agreement between the experimental 
and numerical data, which are very close for both implementations 
of the SA equation.
Special attention should be given for $\alpha= 8^\circ$.
Experimental measurements highlight a natural transition
occurring in the kinematic boundary layer at $32\%$ of the chord length.
In contrast, at $\alpha= 8.5^\circ$ a LSB 
forms near the leading edge, indicating a strong sensitivity to angle of attack in the
around $\alpha= 8^\circ$.
This behavior may also be attributed to an elevated level of free--stream turbulence intensity.
In this specific case, we observed that introducing artificial viscosity
into intermittency transport equation can mitigate the possible 
effect of unmodeled free--stream turbulence assuming that is indeed
not properly accounted.
%
In Fig.~\ref{fig:e387_a8} it is evident that all baseline models predict
laminar separation bubble, which is not present in the experiments.
In contrast, as shown in Fig.~\ref{fig:e387_a8-diff} the SA--R1 model 
(the same trend is confirmed also for other models, not reported here
for brevity) achieves a good agreement with experiments when
an artificial viscosity scaling factor of $c_{art} = 10^{-2}$ is applied.
Note that the required value of $c_{art}$ for SA--SMP model is $10^{-4}$, and $10^{-5}$
 for the $\log \gamma$--SA formulation, confirming the trend previously noted for the SD7003.
This additional dissipation damps spurious LSB and shifts the transition location
to a position consistent with experimental data. 
Fig.~\ref{fig:e387_a12} reports pressure coefficient distribution for 
$\alpha= 12^\circ$. In all the cases, a value of $c_{art} = 10^{-2}$ was found to be
a good choice to suppress pressure oscillations.\\
\hspace{-0.5cm} The force coefﬁcients plots from, Figs.~\ref{fig:e387_cl-alpha-re200k} to~\ref{fig:e387_cl-cd-re300k},
conﬁrm the good correspondence between the numerical and experimental data for both
$\mathrm{Re}$ numbers considered, especially for the lift coefficient.
Specifically, the SA--R1 and SA--SMP model variants yield nearly equivalent drag coefficient predictions, 
while the SA--R1 model shows a reduced overestimation of the lift coefficient in the stall region.
%
%
%
\begin{figure}[htbp]
 \centering
 {\includegraphics[width=0.45\textwidth]{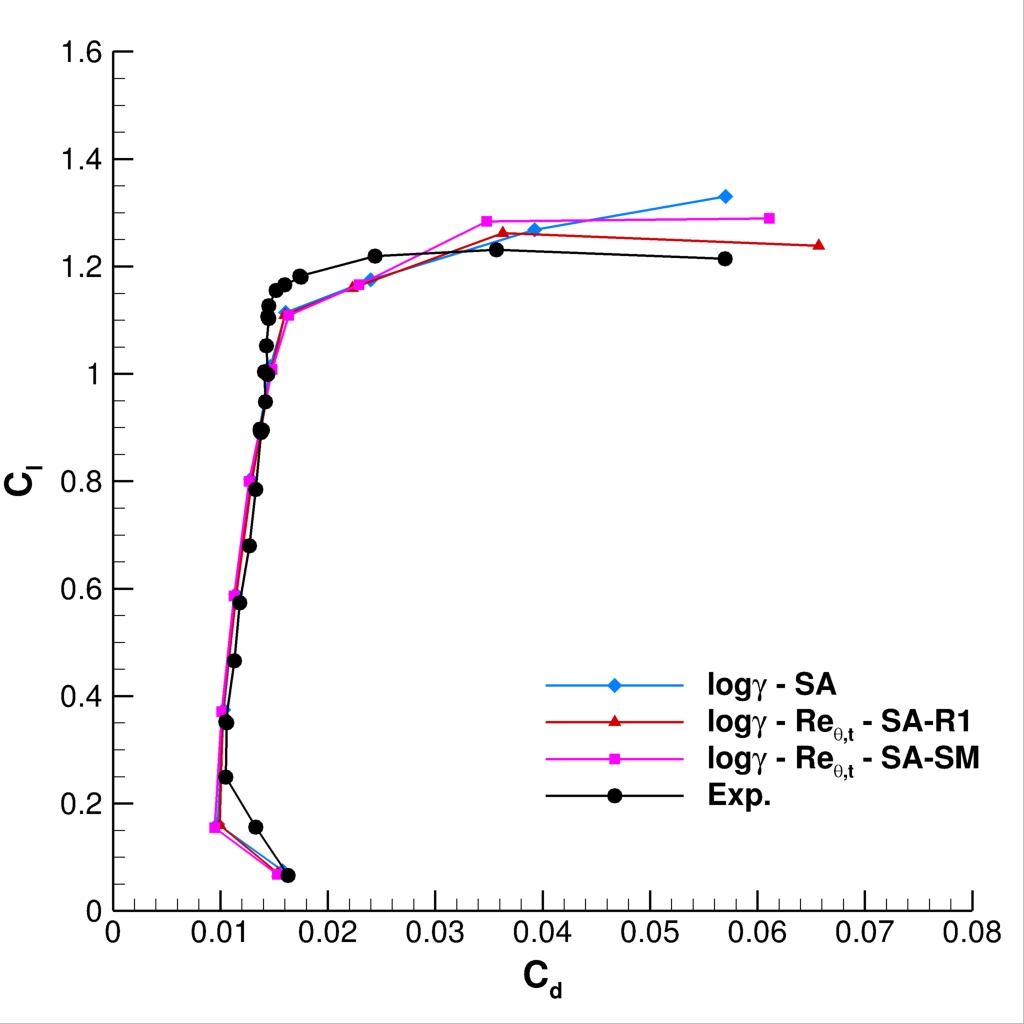}\label{fig:e387_cl-cd-re200k}}
        \caption{E387, $\mathrm{Re} = 2 \cdot 10^5$. Eiffel polar.}
\label{fig:e387_cl-cd-re200k}
\end{figure}
%
Also for this airfoil, the force coefficients data for the highest
angles of attack refer to suitably $c_{art}$ values.
In particular, for the flow models $c_{art} = 10^{-2}$ was employed
for $\alpha \ge 10^\circ$.

\subsection{DU00--W--212}
\begin{figure}[htbp]
 \centering
 {\includegraphics[width=0.45\textwidth]{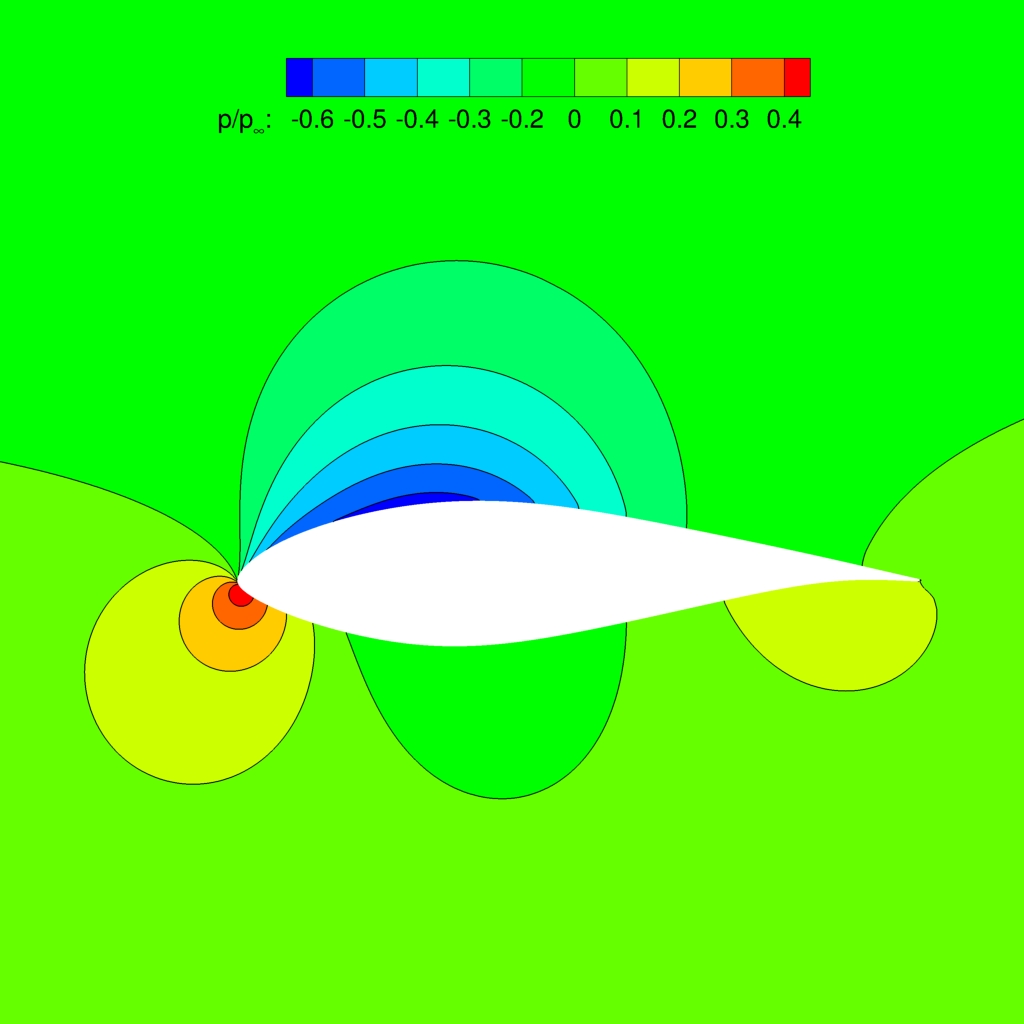}\label{fig:du21}}
        \caption{DU00--W--212, $\mathrm{Re} = 6 \cdot 10^6$, $\alpha = 4^\circ$. Dimensionless pressure field, SA--SM model.}
\label{fig:du21-pressure}
\end{figure}

\begin{figure}[htbp]
 \centering
 {\includegraphics[width=0.45\textwidth]{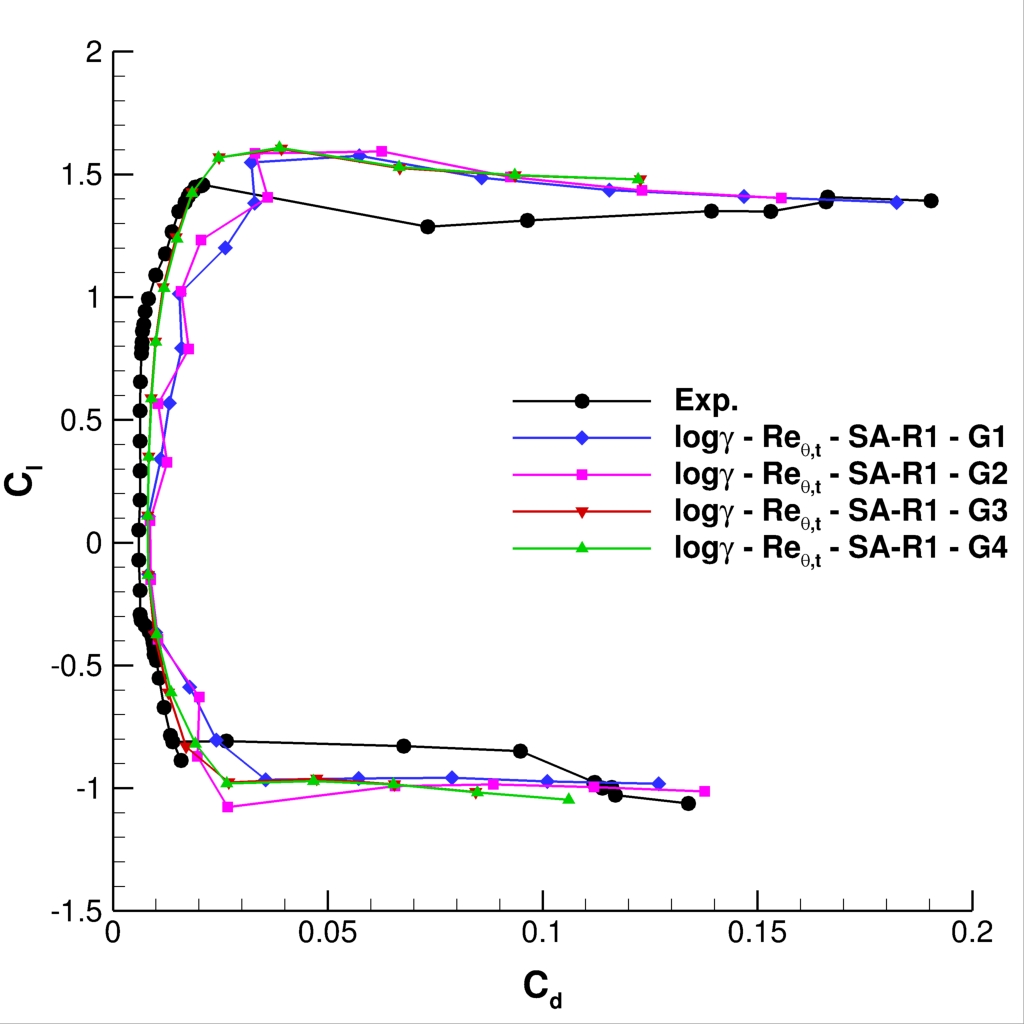}\label{fig:du_cl-alpha-6e6}}
        \caption{DU00--W--212, $\mathrm{Re} = 6\cdot 10^6$. Eiffel polar, grid effect.}
\label{fig:du21_cl-cd-re6e6-grid}
\end{figure}

In this subsection we present some results concerning  
DU00--W--212 airfoil operating a $\mathrm{Re} = 3 \cdot 10^6$ and $\mathrm{Re} = 6 \cdot 10^6$, 
see Fig.~\ref{fig:du21-pressure} for an airfoil representation. 
The free--stream turbulence intensity was set to $0.09 \%$ for $\mathrm{Re} = 3 \cdot 10^6$, while 
a value of $0.2 \%$ was used for $\mathrm{Re} = 6 \cdot 10^6$.
 This case is particularly interesting since it involves a wind turbine airfoil, 
and wind tunnel measurements are available through the AVATAR project~\cite{Ceyan:2015}.\\
Four different O--type grids have been created, named G1, G2, G3 and G4, respectively.
All the grids share the first cell height,  which ensures $O\left(y^+\right) \simeq 1$.
G1 has around $N_c \simeq 1.5 \cdot 10^5$, while G2 has about  $N_c \simeq 4.7 \cdot 10^5$;
in both the cases the far--field boundary is located at approximately $20$ chord lengths away. 
By contrast, G3 contains $10^6$ cells and has its far--field set at 200 chord lengths,
whereas G4 has $1.1\cdot 10^6$ cells with the far--field placed at 300 chord lengths.
It is also important to highlight that G2, G3 and G4 have exactly the same cell 
distribution within the first 20 chord lengths. However, G3 and G4 were specifically 
generated to assess the effect of the far--field boundary distance from the airfoil 
surface on the obtained solution.\\
%
%
%
Fig.~\ref{fig:du21_cl-cd-re6e6-grid} shows the Eiffel polar at $\mathrm{Re} = 6 \cdot 10^6$
obtained using SA--R1 version of the model  with all four computational grids.
It is quite easy to observe that the far--field grid extension significantly
 affects the drag coefficient value. Indeed, G1 and G2 highlight a noticeable
 overestimation of the $C_D$ coefficient. On the other hand, G3 and G4 produce 
indistinguishable results for both the force coefficients. 
For this reason, the results presented hereafter 
are based solely on the G3 grid.\\
%
%
From Fig.~\ref{fig:du_a0} to Fig.~\ref{fig:du_a12-cart}, the pressure coefficient distribution
at $\mathrm{Re} = 3\cdot 10^6$ are shown for angles of attack for $\alpha = 0^\circ, 4^\circ, 8^\circ$ 
and $12^\circ$.
For angles up to $\alpha = 8^\circ$, both the $\log \gamma$--$\Ret$--SA--R1, $\log \gamma$--$\Ret$--SA--SMP 
models exhibit a very good agreement with experimental data. 
Differently, $\log \gamma$--SA model predicts a fully--turbulent flow past the airfoil.
At $\alpha = 12^\circ$, the computed pressure coefficient distribution significantly deviate from the reference data, and
SA--R1 model shows noticeable pressure oscillations near the airfoil leading edge. Nevertheless,
the activation of artificial viscosity, as shown in Fig.~\ref{fig:du_a12-cart}, helps to obtain a physically consistent
solution, although discrepancies with experimental data remain.
\begin{figure}[htbp]
 \centering
 {\includegraphics[width=0.45\textwidth]{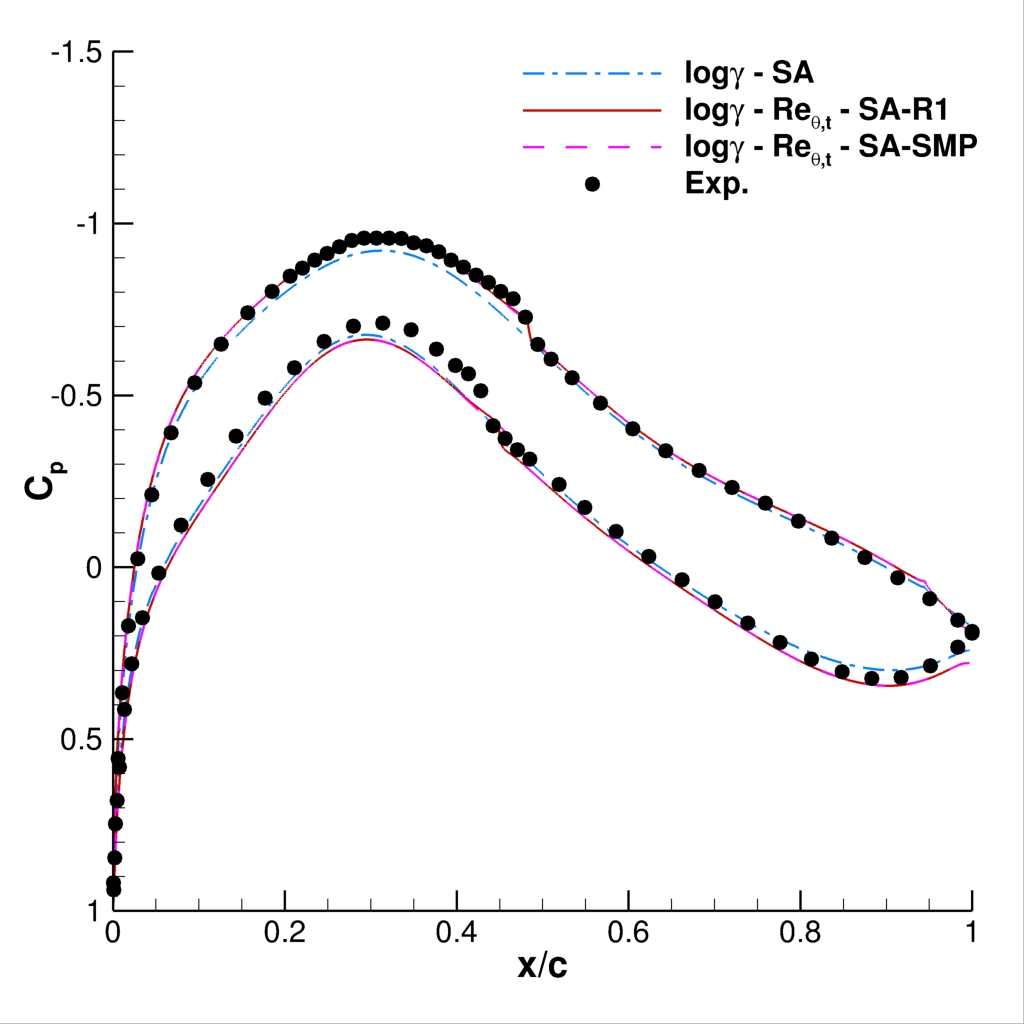}\label{fig:du_a0}}
        \caption{DU00--W--212, $\mathrm{Re} = 3\cdot 10^6$, $\alpha = 0^\circ$. Pressure coefficient distribution.}
\label{fig:du_a0}
\end{figure}

\begin{figure}[htbp]
 \centering
 {\includegraphics[width=0.45\textwidth]{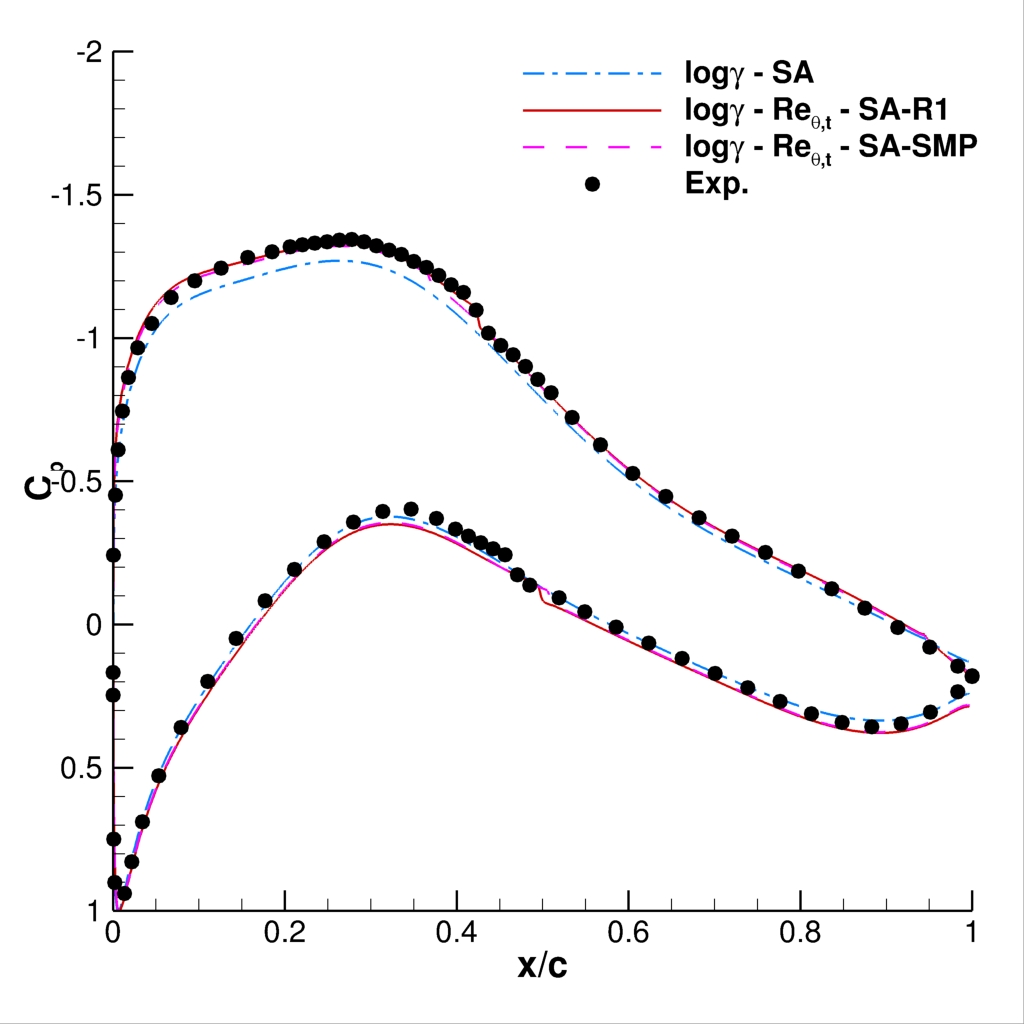}\label{fig:du_a4}}
        \caption{DU00--W--212, $\mathrm{Re} = 3\cdot 10^6$, $\alpha = 4^\circ$. Pressure coefficient distribution.}
\label{fig:du_a4}
\end{figure}

\begin{figure}[htbp]
 \centering
 {\includegraphics[width=0.45\textwidth]{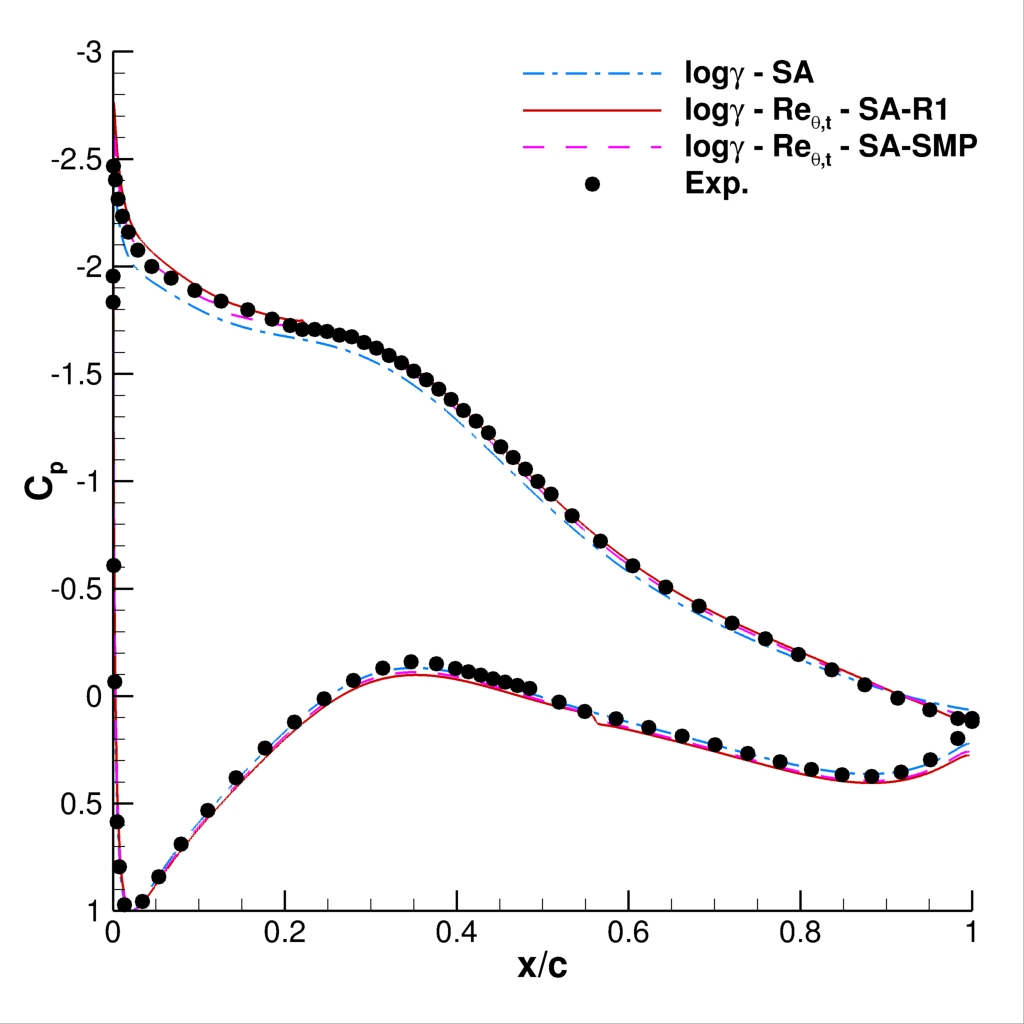}\label{fig:du_a8}}
        \caption{DU00--W--212, $\mathrm{Re} = 3\cdot 10^6$, $\alpha = 8^\circ$. Pressure coefficient distribution.}
\label{fig:du_a8}
\end{figure}

\begin{figure}[htbp]
 \centering
 {\includegraphics[width=0.45\textwidth]{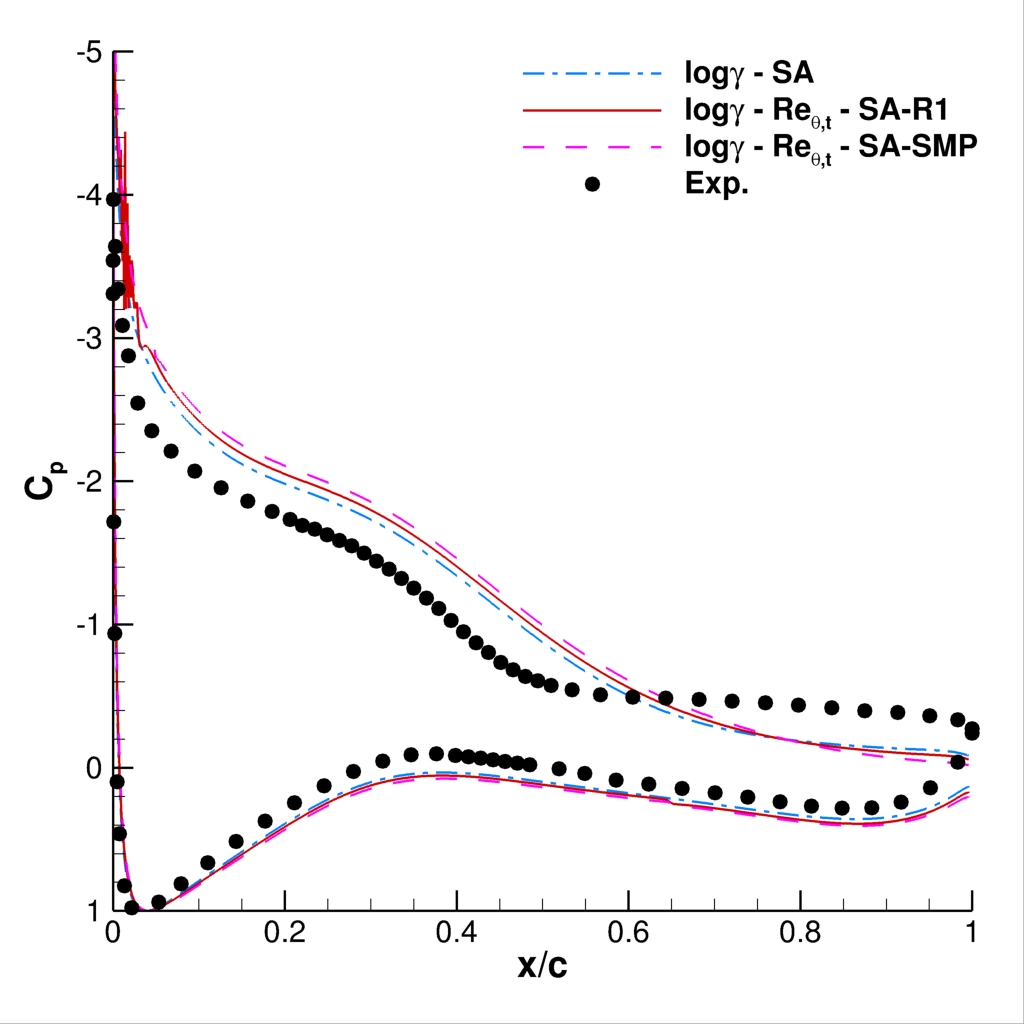}\label{fig:du_a12}}
        \caption{DU00--W--212, $\mathrm{Re} = 3\cdot 10^6$, $\alpha = 12^\circ$. Pressure coefficient distribution.}
\label{fig:du_a12}
\end{figure}

\begin{figure}[htbp]
 \centering
 {\includegraphics[width=0.45\textwidth]{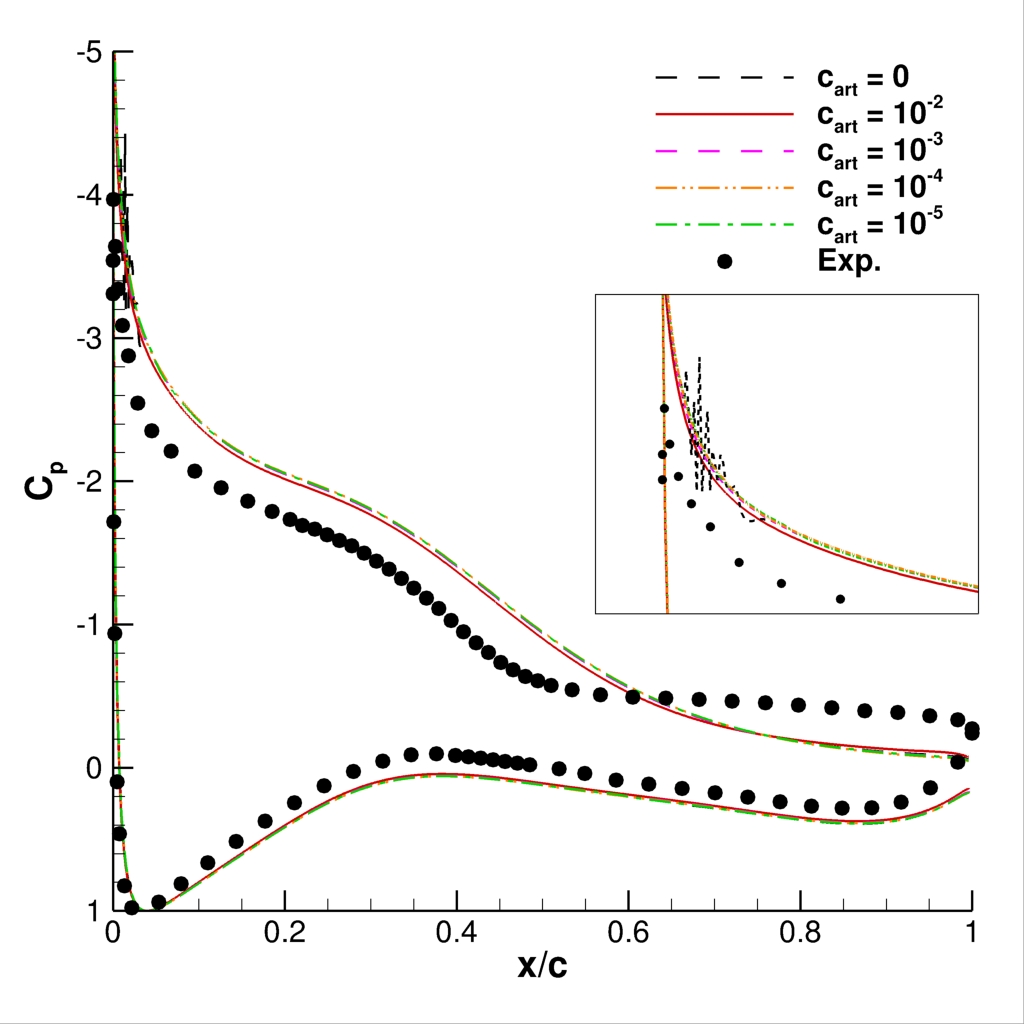}\label{fig:du_a12-cart}}
        \caption{DU00--W--212, $\mathrm{Re} = 3\cdot 10^6$, $\alpha = 12^\circ$. $\log \gamma$--$\Ret$--SA--R1 model,
                 effect of artificial viscosity}
\label{fig:du_a12-cart}
\end{figure}
%
\begin{figure}[htbp]
 \centering
 {\includegraphics[width=0.45\textwidth]{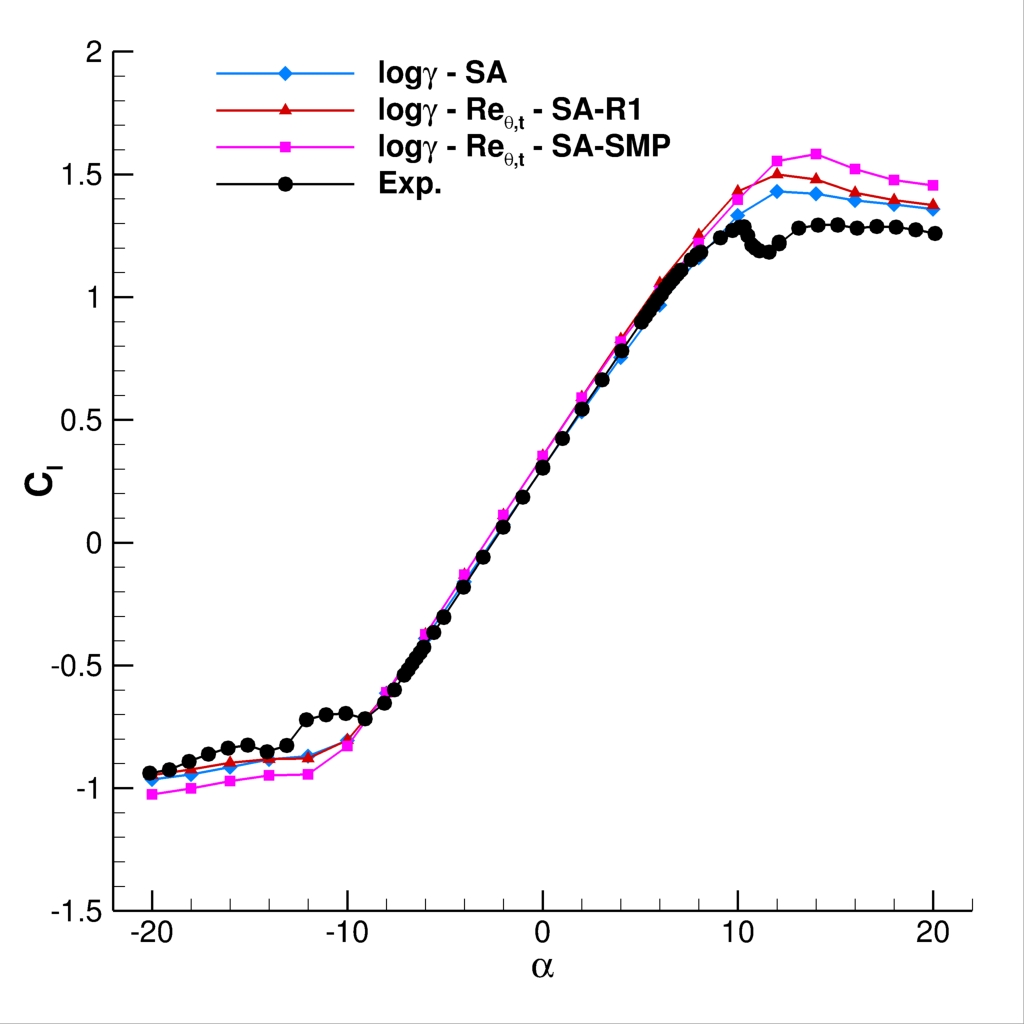}\label{fig:du_cl-alpha-3e6}}
        \caption{DU00--W--212, $\mathrm{Re} = 3\cdot 10^6$. Lift coefficient.}
\label{fig:du_cl-alpha-3e6}
\end{figure}

\begin{figure}[htbp]
 \centering
 {\includegraphics[width=0.45\textwidth]{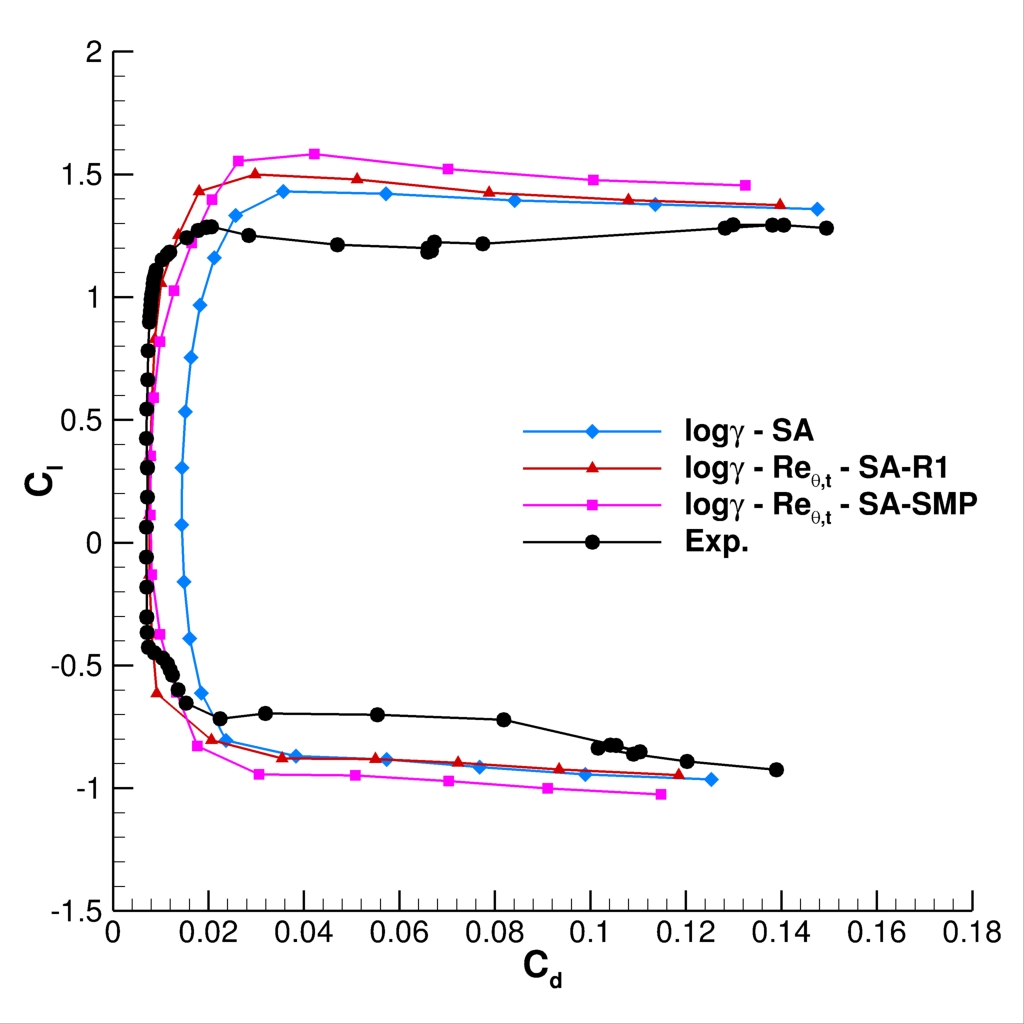}\label{fig:du_cl-cd-6e6}}
        \caption{DU00--W--212, $\mathrm{Re} = 3\cdot 10^6$. Eiffel polar.}
\label{fig:du_cl-cd-3e6}
\end{figure}
Fig.~\ref{fig:du_cl-alpha-3e6} and Fig.~\ref{fig:du_cl-cd-3e6} show the force coefficient plots 
for the case at $\mathrm{Re} = 3 \cdot 10^6$. It is quite clear that the SA--R1 and SA--SMP 
techniques have similar performance in predicting the forces coefficient, 
which are accurately predicted before the stall region but, as is common in RANS--based methods, 
tends to be overestimated after stall. 
In contrast, the $\log \gamma$--SA model version produces satisfactory results in terms lift coefficient.
It systematically overestimates drag, consistently with the observed $C_p$ distribution, where
a  fully turbulent flow is predicted throughout. 
A very similar situation is observed at $\mathrm{Re} = 6 \cdot 10^6$. In this case the one--equation model
is not considered since its poorer performance at lower Reynolds number. 
The lift coefficient results (see Fig.~\ref{fig:du_cl-alpha-6e6}) are very similar across the models up to the stall 
and reproduce very well the experimental measurements.
By contrast, in the stall region SA--R1 model shows a better agreement with the experimental data. 
Regarding the drag coefficient (see Fig.\ref{fig:du_cl-cd-6e6}), the SA-SMP model tends to slightly overestimate 
the reference values, unlike the SA-R1 approach, which yields results closer to the experimental measurements.
%
\begin{figure}[htbp]
 \centering
 {\includegraphics[width=0.45\textwidth]{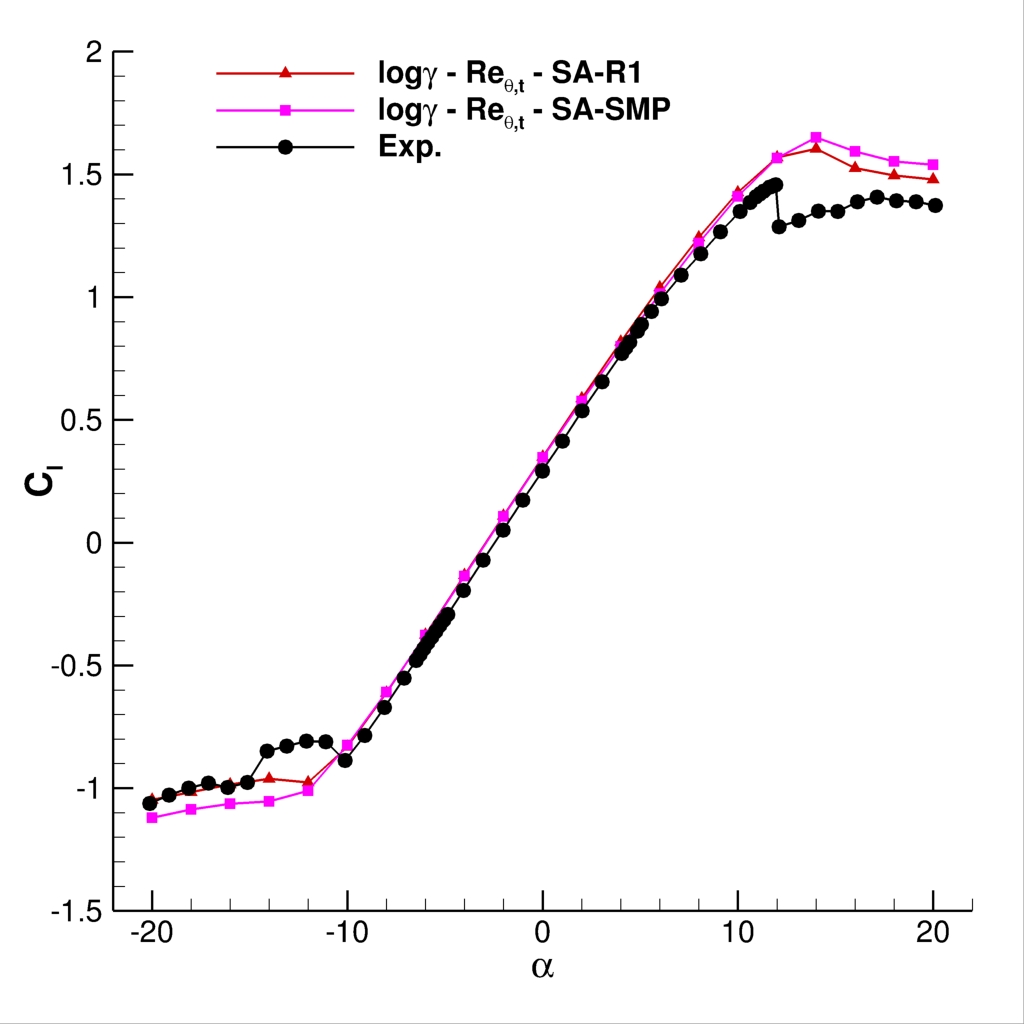}\label{fig:du_cl-alpha-6e6}}
        \caption{DU00--W--212, $\mathrm{Re} = 6\cdot 10^6$. Lift coefficient.}
\label{fig:du_cl-alpha-6e6}
\end{figure}
\begin{figure}[htbp]
 \centering
 {\includegraphics[width=0.45\textwidth]{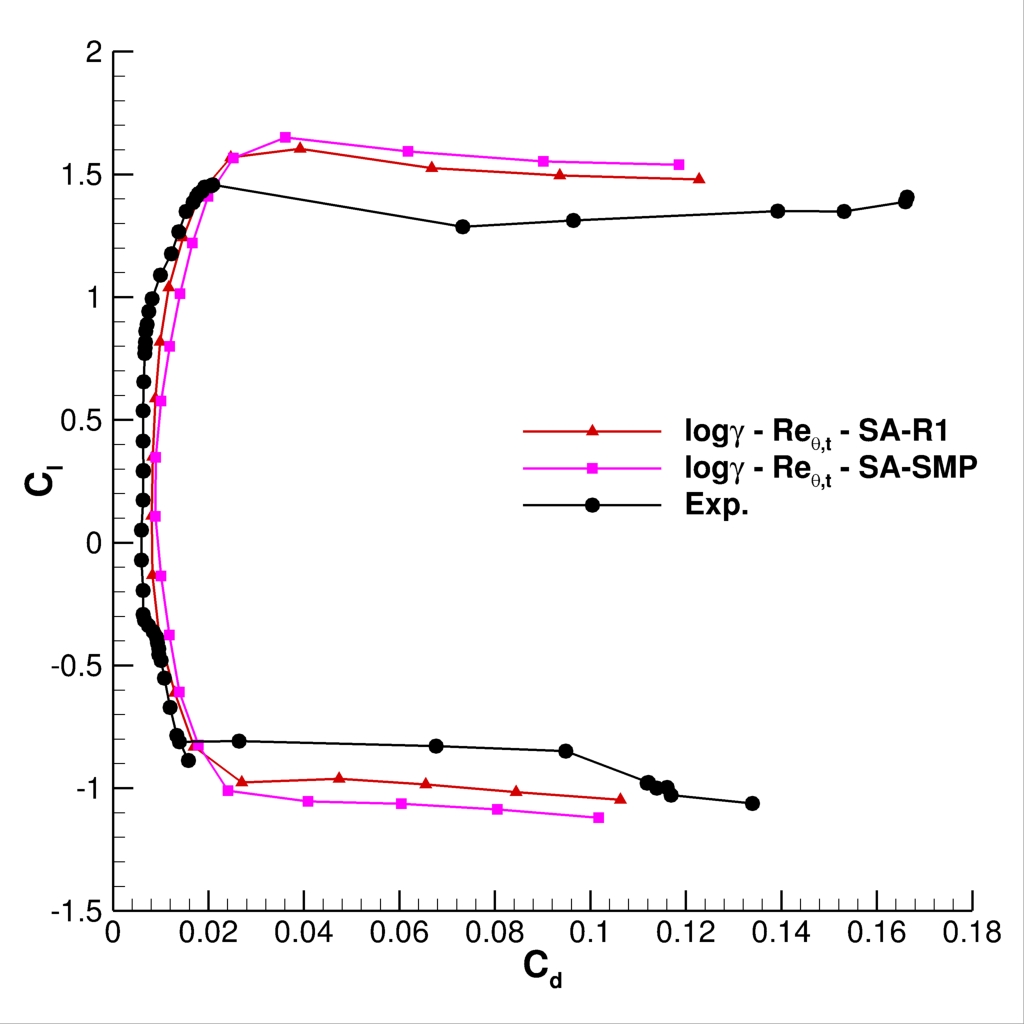}\label{fig:du_cl-cd-6e6}}
        \caption{DU00--W--212, $\mathrm{Re} = 6\cdot 10^6$. Eiffel polar.}
\label{fig:du_cl-cd-6e6}
\end{figure}
Finally, we want to emphasize that, for this airfoil, the force coefficients data are 
not significantly affected by artificial viscosity when used.

\section{Conclusions}\label{sec:concl}
This paper presents the latest developments of RANS local--correlation based 
laminar--to--turbulent transition methods built upon the Spalart--Allmaras turbulence model.
As noted above, this specific formulation is particularly appealing because 
it relies on a single turbulence equation, making it highly suitable 
for flow problems involving a large number of grid cells.\\
This paper focuses on proposing and testing a new stabilized formulation for intermittency
transport equation designed to ensure robustness and reliability across a wide range of Reynolds numbers.
First of all, the intermittency equation is re--written in logarithmic form
in order to enforce positive values for the primitive variable.
A further stabilization procedure, based on the limiting 
of the artificial integrated energy associated with logarithmic intermittency, 
is also introduced and discussed.
This approach helps to prevent excessively high intermittency values, which can act as potential 
triggers for floating-point exceptions.
The model formulation proved to be numerically stable across several flow benchmarks 
within a relevant range of Reynolds numbers. 
The last numerical ingredient introduced is a gradient driven artificial viscosity for the logarithmic
intermittency equation aimed at damping Gibbs--like phenomena observed in the pressure within the transition region.\\
Lastly, the above discussed techniques were successfully applied to both $\gamma$ and $\gamma$--$\Ret$
models by also employing two different formulations for the SA equation source terms and named SA--R1 and
SA--SMP, respectively.\\
Several well-established flow benchmarks have been presented, and overall, the results indicate 
that $\log \gamma$--$\Ret$--SA models provide consistent accuracy for all the considered 
Reynolds numbers.
The only case where the $\log \gamma$--$\Ret$--SA--SMP is clearly worse than the other two--equation version is the DU00--W--212 at  $\mathrm{Re} = 6 \cdot 10^6$.
By contrast, the model using only one--equation for transition produces good results only at medium
low Re numbers as clearly shown.\\
A final remark concerns for the artificial viscosity. 
%
Notably, this stabilization mechanism proved effective in all the considered test cases
and transition models, although different levels of artificial diffusion
were required to achieve smooth solutions.
In particular, our computations put in evidence that 
SMP variant of the SA equation,  as well as
the one--equation transition model, tend to introduce a higher degree 
of numerical diffusion compared to the R1 formulation.\\
It is worth noting that without the proposed stabilization techniques, 
the solution process becomes unreliable or completely fails. 
However, the new formulation ensures robustness across a broad Reynolds number range.
This elements makes our models particularly suitable for practical applications, 
especially where accurate transitional modeling is needed within industrial RANS frameworks.
Finally, note also that although the method has been extensively tested on the Spalart--Allmaras model coupled with the 
$\gamma$ and $\Ret$ equations, the same stabilization procedure can be straightforwardly 
extended to the $k$–$\omega$ model.\\
Future work will focus on potential Detached--Eddy Simulation (DES) extensions 
to enhance the model's performance in the stall region.

\section{Acknowledgements}
%
%
The authors would like to express their gratitude  Pietro Catalano and Donato de Rosa 
from CIRA for kindly providing the SD7003 airfoil grid as well as the related reference data.

\section*{Data availability}
The data that support the findings of this study are available from the corresponding author upon reasonable request.

\section{References}
\bibliography{aipsamp}

\end{document}